\documentclass{emulateapj}
\usepackage{graphicx}

\slugcomment{Accepted by Astrophysical Journal.}

\shorttitle{Spectral Energy Distribution of z=2.1 LAEs}
\shortauthors{Guaita et al.}

\begin{document}

\title{Ly$\alpha$-Emitting Galaxies at z = 2.1: Stellar Masses, Dust and Star Formation Histories from Spectral Energy Distribution Fitting\footnote{Based on observations obtained at Cerro Tololo Inter-American Observatory, a division of the National Optical Astronomy Observatory, which is operated by the Association of Universities for Research in Astronomy, Inc., under cooperative agreement with the National Science Foundation.}}

\author{Lucia Guaita\altaffilmark{1,}\altaffilmark{2},}
\affil{Departmento de Astronom\'\i a y Astrof\'\i sica, Universidad Cat\'olica de Chile,
    Santiago, Chile}
\affil{Department of Astronomy, Oscar Klein Center, Stockholm University,  AlbaNova, Stockholm SE-10691, Sweden}    
\email{lguai@astro.su.se}    

\author{Viviana Acquaviva\altaffilmark{3}}

\author{Nelson Padilla\altaffilmark{1},Eric Gawiser\altaffilmark{3}}

\author{Nick Bond \altaffilmark{3}, R Ciardullo\altaffilmark{4}, Ezequiel Treister\altaffilmark{5,}\altaffilmark{6,}\altaffilmark{11}}

\author{Peter Kurczynski\altaffilmark{3}, Caryl Gronwall\altaffilmark{4}, Paulina Lira\altaffilmark{7}, and Kevin Schawinski\altaffilmark{8,}\altaffilmark{9,}\altaffilmark{10}} 

\altaffiltext{1}{Departmento de Astronom\'\i a y Astrof\'\i sica, Universidad Cat\'olica de Chile,
    Santiago, Chile}
\altaffiltext{2}{Department of Astronomy, Oscar Klein Center, Stockholm University,  AlbaNova, Stockholm SE-10691, Sweden}    
\altaffiltext{3}{Rutgers, The State University of New Jersey, NJ 08854}
\altaffiltext{4}{Department of Astronomy \& Astrophysics Penn State University, State College, PA 16802 }
\altaffiltext{5}{Institute for Astronomy, Honolulu, Hawaii 96822-1839}
\altaffiltext{6}{Institute for Astronomy, 2680 Woodlawn Drive, University of Hawaii, Honolulu, HI 96822}
\altaffiltext{7}{Departamento de Astronom'a de la Universidad de Chile, Santiago, Chile}
\altaffiltext{8}{Einstein Fellow}
\altaffiltext{9}{Department of Astronomy, Yale University, New Haven, CT 06511}
\altaffiltext{10}{Yale Center for Astronomy and Astrophysics, Yale University, P.O. Box 208121, New Haven, CT 06520}
\altaffiltext{11}{Departmento de Astronom\'\i a, Universidad de Concepci\'on, Concepci\'on, Chile}

\begin{abstract}

We study the physical properties of 216 $z\simeq2.1$ LAEs discovered in an ultra-deep narrow-band MUSYC image of the ECDF-S. We fit their stacked Spectral Energy Distribution (SED) using Charlot \& Bruzual templates. We consider star formation histories parametrized by the e-folding time parameter $\tau$, allowing for exponentially decreasing ($\tau>0$), exponentially increasing ($\tau<0$), and constant star formation rates.  We estimated the average flux at 5015 {\AA} of our LAE sample, finding a 
non-detection, which translates into negligible HeII line emission at $z\simeq2.1$. In addition to this, the lack of high equivalent width Ly$\alpha$ line objects ruled out the hypothesis of a top-heavy IMF in LAEs.  The typical LAEs of our sample are characterized by best fit parameters and 68\% confidence intervals of log(M$_*$/M$_{\odot}$)=8.6[8.4-9.1], E(B-V)=0.22[0.00-0.31], $\tau$=-0.02[(-4)-18] Gyr, and age$_{\rm SF}$=0.018[0.009-3] Gyr. Thus, we obtain robust measurements of low stellar mass and dust content, but we cannot place meaningful constraints on the age or star formation history of the LAEs. We also calculate the instantaneous SFR to be 35[0.003-170]  M$_{\odot}$ yr$^{-1}$, with its average over the last 100 Myr before observation giving $<$SFR$>_{100}$=4[2-30] M$_{\odot}$ yr$^{-1}$.  When we compare the results for the same star formation history, typical LAEs at $z\simeq2.1$ appear dustier and show higher instantaneous SFRs than $z\simeq3.1$ LAEs, while the observed stellar masses of the two samples seem consistent.  Because the majority are low-mass galaxies, our typical LAEs appear to occupy the low-mass end of the distribution of star forming galaxies at $z\sim2$.  We perform SED fitting on several sub-samples selected based on photometric properties and find that LAE sub-samples at $z\simeq2.1$ exhibit heterogeneous properties. The typical IRAC-bright, UV-bright and red LAEs have the largest stellar mass and dust reddening. The typical UV-faint, IRAC-faint, and high equivalent width LAE sub-samples appear less massive  ($<10^9$ M$_{\odot}$) and less dusty, with E(B-V) consistent with zero.

\end{abstract}


\section{INTRODUCTION}

Lyman Alpha Emitting (LAE) galaxies have been observed at very high redshift, thanks to their strong emission lines enabling discovery even without continuum detection. Many studies have been carried out to investigate their nature, using photometric and spectroscopic data (e.g., \citealt{CHu1998}, \citealt{MR:2002}, \citealt{Ouchi:2003}, \citealt{Fin:2008},  \citealt{Ouchi:2008}, \citealt{Gronwall:2007}, \citealt{Nilsson:2007}, \citealt{Nilsson:2009}).
A description of the physical nature of LAEs can be obtained from spectral energy distribution (SED) analysis (\citealt{Gawiser2009}, 
\citealt{Walcher2010} for reviews and \citealt{KannappanGawiser07} for caveats).
Multi-wavelength surveys allow us to study the SEDs, 
which can be obtained from individual object flux densities, when there is enough signal-to-noise, or through photometric 
stacking of galaxy samples. At $z\ge3$ LAEs are young, nearly dust-free, low-mass objects ($z\simeq3.1$ \citealt{Gawiser:2006b}, \citealt{Nilsson:2007}, \citealt{Gawiser:2007}, \citealt{Lai:2008}, \citealt{Ono2010}, $z\simeq4.5$ \citealt{Fin:2008}, $z\simeq5$ \citealt{Yuma:2010}, $z\simeq6$ \citealt{Pirzkal:2007}, \citealt{Ono2010b}, \citealt{Ouchi:2010}). 
Furthemore, Gawiser et al. (2007) found that the typical LAEs at $z\simeq3.1$ are characterized by a median starburst age of $\sim20$ Myr, a low stellar mass of M $\sim10^9$ M$_{\odot}$, modest SFR values around $\sim2$ M$_{\odot}$ yr$^{-1}$, and a negligible amount of dust (A$_V \leq$ 0.2), indicating that they are galaxies in the early phases of a burst of star formation (SF). They also measured the spatial clustering 
 of the LAE sample, and inferred that  LAEs are hosted by the population of dark matter halos with masses greater than 
$ \log({\rm M}_{\rm min}$/M$_{\odot}$) = 10.6$^{+0.5}_{-0.9}$. 
From the evolution of dark matter mass with redshift (\citealt{Shetormen:1999}, \citealt{Francke:2008}), 
their results imply that $z\simeq3.1$ LAEs 
can evolve into Milky Way-type galaxies at $z\sim0$ (\citealt{Gawiser:2007}).
This adds to evidence from SED fitting and rest-frame ultraviolet sizes that high-redshift LAEs could be building blocks of more evolved lower redshift galaxies (\citealt{Pascarelle1998}, \citealt{Ouchi:2003}, \citealt{Ouchi:2010}). 

LAEs can therefore occupy the low-mass end of high-z galaxy populations, where the bulk of the galaxies lie. Other well-studied populations, like Lyman Break Galaxies (LBG) or Sub-Millimeter Galaxies (SMG), represent the rare luminous tail of the high-redshift luminosity function. However, despite having low mass, LAEs do not have particularly high number density, implying that they are selected from the low-mass bulk of high-redshift galaxies due to special conditions in which the Ly$\alpha$ photons are produced and are able to escape. These conditions involve star formation, responsible for the ionizing photon production, the spatial and velocity distribution of neutral hydrogen in the interstellar medium (ISM) and the ISM dust geometry (\citealt{V:2006} as an example). Large samples are crucial to better constrain typical properties of star forming galaxies in general, and LAEs in particular. The comparison of LAE properties with those of color-selected star forming galaxies has shown that Ly$\alpha$ emission can also be detected in older, more massive, dusty galaxies (\citealt{Shapley:2001}, \citealt{Tapken:2007}, \citealt{Pentericci:2009}, \citealt{kornei2010}). 

Recently  \citeauthor{Nilsson:2009} (2009) claimed that at lower redshift ($z\simeq2.3$), LAEs present more diversity in photometric properties and hence a possible evolution from $z\ge3$ to $z\simeq2$, in 1 Gyr of Universe life. This is confirmed in their recent SED analysis (Nilsson et al. 2011).
\citeauthor{Guaita2010} (2010, hereafter Gu10) obtained a sample of 250 LAEs at $z\simeq2.1$ in the Extended Chandra Deep Field - South (ECDF-S), one of the fields of MUSYC (Multi-wavelength Survey by Yale-Chile, \citealt{Gawiser:2006a}\footnote{http://physics.rutgers.edu/$\sim$gawiser/MUSYC}).  
The clustering properties of this sample were similar to LAEs at $z\simeq3.1$, implying that they also serve as building blocks of present-day $L^*$ galaxies. The Ly$\alpha$ luminosity and rest-frame equivalent width (EW) limits (log(L$_{Ly\alpha}$) = 41.8, EW$_{rest-frame}=20$ {\AA}) adopted in Gu10 were the same  
as those at $z\simeq3.1$ studied by Gronwall et al. (2007). This allows us to make an unbiased comparison between the photometric properties (Gu10) and spectral energy distributions (this work) of the two LAE sample, investigating the LAE evolution claimed by Nilsson et al. (2009, 2011).

Section \S\ref{sec:sample} of this paper describes the LAE sample and sub-samples based on rest-frame NIR flux density, the rest-frame UV slope, represented by the $B$-$R$ color, rest-frame UV 
magnitude, equivalent width.
In Section \S\ref{sec:model}, we describe the stellar population synthesis templates used in the SED fitting. 
We present the results in Section \S\ref{sec:results}, including the comparison with SED results at $z\simeq3.1$. 
In Section \S\ref{sec:discussion} we discuss the implications of the results of the SED fitting of our sample of $z\simeq2.1$ LAEs and its sub-samples. 
We summarize the inferred physical properties of z=2.1 LAEs in Section \S\ref{sec:conclusion}.

We assume a $\Lambda$CDM cosmology consistent with WMAP 5-year results (\citealt{Dunkley:2009}, their Tab. 2), adopting the mean parameters  $\Omega_m=0.26$, $\Omega_{\Lambda}=0.74$, H$_0=70$ km~s$^{-1}$~Mpc$^{-1}$, $\sigma_8=0.8$.

\section{THE SAMPLE}
\label{sec:sample}

\subsection{Stacked Multi-wavelength Photometry}
\label{sec:stack}

We used the sample of 250 LAEs found in the 
deep narrow-band image of ECDF-S by Gu10
at $z\simeq2.1$.  
The ten criteria used in the selection of Ly$\alpha$ emitting galaxies are listed in \S4 of Gu10.
Morphology analysis of the 250 $z\simeq2.1$ LAEs (\citealt{Bond2011}) revealed the presence of 34 extended objects with sizes larger than $\sim$ 0.6$''$ 
in the HST/ACS $V$ broad band (point spread function of 0.1$''$). This size corresponds to a half-light radius of 2.5 kpc at $z\simeq2.1$.
Previous studies on the morphology of Ly$\alpha$ emitting galaxies helped us in deciding to exclude these objects in this work.
\citealt{Bond2009} also analyzed morphologically a sample of $z\simeq3.1$ LAEs, using HST broad band images. They found that the typical high redshift LAE was barely resolved at the $\sim0.1''$ angular resolution of the HST/ACS $V$-band image and that very few were larger than 1.5 kpc. 
In Bond et al. (2011), some extended sizes were found for $1.5<z<3$ star forming galaxies, but objects with half-light radius bigger than $2-3$ kpc were very rare at $z\sim2$. The median half-light radius of $z\simeq2.1$ LAEs is, instead, 1 kpc and the majority of the LAEs are characterized by sizes smaller than 2 Kpc. 

These extended objects, if real LAEs, must constitute a separate population not seen at $z\simeq3.1$ and are worth studying separately. Not yet having spectroscopy of them to verify their redshift, we prefer the conservative approach of excluding them from SED analysis. 
The majority is characterized by low Ly$\alpha$ luminosity and limited equivalent width.  
Excluding them from the sample, the photometric results presented in Gu10 are still valid. This is mainly due to the fact that we had reported median typical properties and also because number density, equivalent width, luminosity distributions and star formation rate estimations were derived for the half of the sample complete in terms of EW. 
The clustering results obtained for the sample of 216 objects is also consistent with those published in Gu10 within the errors.
The low-EW, low-Ly$\alpha$ luminosity part of the sample is surely composed by real LAEs, but it is also the most likely to be affected by low-redshift interlopers. A careful inspection of ground and space-based data is needed, but spectroscopy is the most efficient way to disentangle their nature. 
The median stacking approach from the observed optical to IR we adopt below (see \S \ref{sec:median} for a discussion of the use of median statistic) should minimize the influence of contaminants unless they are numerous, and we indeed observed that the SEDs of the full sample and of the sub-samples do not change significantly upon removing these extended objects. However, we removed them from the full sample since we do not think they are representative of the typical LAE at $z\simeq2.1$. Our final $z\simeq2.1$ LAE sample is thus composed of 216 objects.

\begin{figure}[!ht]
\begin{center}
\includegraphics[width=54mm, height=40mm]{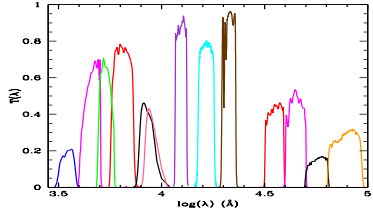}
\caption{Filter transmission curves for MUSYC ECDF-S broadband photometry. From left to right: $U, B, V, R, I, z, J, H, K$, IRAC 3.6, 4.5, 5.8, 8.0$\mu$m.}
\label{filters}
\end{center}
\end{figure}

Using the SExtractor (\citealt{bertin1996}) program in dual image mode, we estimated the aperture fluxes of the narrow-band detected galaxies in all the broad bands (Tab. \ref{tab:limitingMAG}) of the MUSYC survey shown in Fig. \ref{filters}.  These included the $U, B, V, R, I, z$ optical \citep{Cardamone:2010}, $J, H, K$ near-infrared \citep{Taylor:2009}, and IRAC 3.6, 4.5, 5.8, 8.0 $\micron$ infrared bands \citep{Damen2010}.
The apertures in each band were optimized by calculating the $curve~of~growth$ of bright point sources. This led us to choose 1.2$''$ diameter apertures for all the optical bands, 1.6$''$ for $J$, 1.8$''$ for $H$, 2.0$''$ for $K$, 1.6$''$ for 3.6 $\micron$, 1.8$''$ for 4.5 $\micron$, 2.0$''$ for 5.8 $\micron$, and 2.2$''$ for 8.0 $\micron$. 
We used the fractional signal inside the optimal apertures for point sources to transform the aperture fluxes to total fluxes. This assumed that $z\simeq2.1$ LAEs were unresolved in $\sim$ 1$''$ seeing, which our morphology analysis showed to be an excellent approximation. To test the accuracy, we looked for matches between our LAE sample and version2 of the SIMPLE IRAC catalog in the ECDF-S (\citealt{Damen2010}), composed by $\sim$ 45000 objects. 
For the 85 common objects  (34\% of our sample) we could compare the SIMPLE photometry with ours. The median fluxes agreed well with a maximum difference of 15\% in the 8.0 $\micron$ band image, which is characterized by the worst resolution.

We are interested in recovering the $typical$ observed SED of $z\simeq2.1$ LAE population. We stacked the LAE measured fluxes (including formal non-detections and negative fluxes) by computing the median flux in each band. 
Some objects were excluded from the stacking in the NIR and IR bands. The NIR MUSYC images cover a slightly smaller area on the sky than the optical ones, and in the NIR flux stacks we excluded from the 216 LAE sample the galaxies outside this NIR area (3 objects were excluded from the flux stacks in $J$ and 47 in $H$ band). 
In the IRAC 3.6 $\mu$m image, we selected ``clean'' regions, where the photometry of the objects was not affected by neighbors; we also excluded the galaxies with one or more than one IRAC detected objects in a searching radius greater than 1$''$, but smaller than 4$''$ ($\sim$ twice the IRAC point spread function), and 3 objects that were not properly de-blended. The IRAC fluxes from all 4 bands were then obtained from the median among 114 objects. The LAEs' own properties played no role in the classification of ``clean'' regions, so the resulting photometry should be unbiased.  

The errors on the stacked fluxes were estimated as follows. For each data point, there are two sources of uncertainty: the photometric error on the measurement and the sample variance. The former was determined by the quality of the observations and the instrumentation, while the latter quantified the error introduced by assuming that the properties inferred from one sample were a good representation of the general properties of LAEs at this redshift in the Universe as a whole. We took sample variance into account by means of the bootstrapping technique (B. Efron, 1979, $Bootstrap ~ Methods$). By Monte-Carlo sampling with replacement our catalog of 216 LAEs, we built a large number of catalogs of the same size as the original and we computed the sample variance error as the scatter in the median fluxes in each band. This technique was robust when the number of realizations was large enough that the estimation of the mean of the median fluxes approached the flux in the ``real" catalog and when the scatter did not change if the number of realizations was increased. Our choice of using 100,000 Monte-Carlo realizations amply met both these requirements. 
We ignored the formal aperture flux uncertainties while doing the bootstrap, where sample variance dominated. 
The zero-point uncertainties were then added in quadrature to the sample variance error. The zero-point uncertainty was 5\% for all the bands, but 10\% for $U$ and $H$ bands, due to the known difficulties in calibrating a blue filter and because the $H$-band image is a combination of smaller fields which do not have exactly the same photometric zero-point. 
The disagreement between ours and SIMPLE photometry could be due to a zero-point offset, so we were taking into account that with this additional 5\% uncertainty. The observed (N)IR bands were characterized by higher total uncertainty than those observed in the optical bands, due primarily to sample variance between the bootstrap samples. On average the sample variance contribution to the total error budget was $\sim$ 80\%.

\subsection{Sub-samples of LAEs}
\label{sec:subsamples}
 
We are interested in the $typical$ SED of the whole sample of LAEs at $z\simeq2.1$, but also in possibly detecting diverse LAEs populations within the sample. Lacking sufficient signal-to-noise to fed the SEDs of most individual objects, we separated the sample into sub-samples based on photometric properties and stacked their fluxes following the method described in \S \ref{sec:stack} and then in \S \ref{sec:median}. We fit the SEDs of the following sub-samples, whose specifics are listed in Tab. \ref{tab:subsampledefinition}.

1) The full sample of 216 $z\simeq2.1$ LAEs (FULLsample hereafter).

2) We split the full sample into two based on the 3.6 $\mu$m flux density. Lai et al. (2008) separated the $z\simeq3.1$ LAEs in two sub-samples, based on the IRAC 3.6 $\micron$ 2$\sigma$ detection (f$_{3.6 \micron}\geq0.3 \mu$Jy) and non-detection. 
They observed that the IRAC-detected sub-sample (L08det hereafter) 
 was composed by older (160 Myr) and more massive (10$^{10}$ M$_{\odot}$) LAEs than the IRAC-undetected (L08und hereafter) objects. 
Our flux cut at f$_{3.6 \micron}=0.57 \mu$Jy corresponds to the same luminosity cut, 
blue-shifted to $z\simeq2.1$, allowing direct comparison with their results. Only objects in ``clean'' IRAC regions were considered for these samples; 
the resulting IRAC-bright sub-sample contains 47 LAEs and the IRAC-faint sub-sample contains 67 LAEs. The separation based upon the 3.6 $\mu$m flux was made just among the objects located in the IRAC ``clean'' regions. In the ``dirty'' regions the 3.6 $\mu$m flux estimation is not reliable due to effect of neighboring sources; this did not affect the fluxes for any of the non-IRAC bands, which were accepted for the analysis on the full field.

3) Fig. 6a of Gu10 plotted the observed $R$ band (rest-frame UV) magnitude versus the $B$-$R$ color (proportional to the rest-frame UV slope) 
and observed a red branch at  $R<23$. We are interested in fitting these galaxies' SED to infer their nature. Therefore we identified a sub-sample on the basis of $B-R$ color. Approximately 15\% of the full sample (34 LAEs) is characterized by $B-R\geq0.5$, red-LAEs hereafter.   

4) We defined two sub-samples of LAEs split at $R=25.5$. The continuum-bright (UV-bright hereafter) sub-sample 
of 118 LAEs enables a direct comparison with the SED parameters of $R<25.5$ ``BX'', star forming galaxies in the same range of redshift (\citealt{Steidel:2004}).
The remaining 98 LAEs are classified as UV-faint.  

5) We used the estimations of L$_{Ly\alpha}$ and EW from Gu10 to define a complete LAE sample of 119 objects with L$_{Ly\alpha}\geq10^{42.1}$ erg sec$^{-1}$. We investigate the SED dependence on equivalent width, dividing the complete sample at the median EW$_{rest-frame}$ of 66 {\AA}, obtaining 60 high-EW (EW$_{rest-frame}$~$\geq66$ {\AA}) and 59 low-EW (EW$_{rest-frame}$~$<66$ {\AA}) sub-samples. 

We investigated the possible overlaps among the sub-samples in Tab. \ref{tab:overlap}. 
Just 34\% (16/47) of the IRAC-bright galaxies are also red LAEs, but 89\% (42/47) of them are also UV-bright. Among the 118 UV-bright objects, 42/118 (36\%) are IRAC-bright, 
16/118 (13\%) are IRAC-faint, and the remaining 60/118 (51\%) are not included in either IRAC sample due to nearby neighbors.  
Around 50\% of the UV-faint objects also belong to the sub-sample of IRAC-faint. 
Therefore we expect some similarities between IRAC-bright and UV-bright and between IRAC-faint and UV-faint SED results, but none of the sub-samples have sufficient overlap to be effectively identical. One exception are the blue-LAEs, which by virtue of containing $\sim$ 85\% of the total sample also contain the majority of most sub-samples. We expect results from blue-LAEs to be very similar to those for the FULLsample.

\subsection{BX galaxy comparison sample}
\label{sec:BX}

As a means of comparing the SED properties of star forming galaxies at $z\simeq2$, we fit the SED of a sample of spectroscopically confirmed color-selected (``BX'' hereafter) galaxies found in the MUSYC ECDF-S catalog. We selected these galaxies using a two-color diagram including $U-V$, $V-R$ colors and $R$ band magnitude, as in Steidel et al. (2004). The criteria took advantage of the decrement in the spectrum due to neutral Hydrogen absorption at wavelengths just bluer than Ly$\alpha$. We first generated rest-frame spectra of star forming galaxies using the \citealt{BruzualCharlot2003} code at ages between 2.5 Myr and 3 Gyr. We then added a reddening contribution parametrized by \citealt{Calzetti2000} law. We evolved the resulting spectra with redshift and calculated the $U-V$, $V-R$ colors at every redshift. The reddening amount was the dominant parameter in widering the $V-R$ range for star forming galaxies. 
For our filter curve configuration, the $U-V$, $V-R$ space that selected redshifts in the range $2.0<z<2.7$ was defined by 
\begin{equation}
V-R\geq-0.15
\label{bxcriteria2}
\end{equation}
\begin{equation}
U-V\geq3.45(V-R)+0.32
\label{bxcriteria3}
\end{equation}
\begin{equation}
U-V<4.46(V-R)+1.6
\label{bxcriteria4}
\end{equation}
\begin{equation}
V-R\geq0.03(U-V)+0.24.
\label{bxcriteria5}
\end{equation}
We also asked for magnitudes $23<R<25.5$. By definition (\citealt{Steidel:2004}) BX galaxies are bright in the continuum; the $R>23$ requirement allowed us to exclude galactic stars, which are located in the same region of the two-color diagram (Gu10 Fig. 7b).
Just like Lyman Break galaxies at $z\sim3$, BX galaxies could show Ly$\alpha$ emission line in their spectra, depending on their interstellar medium properties.
Among the $\sim$ 84000 sources in the MUSYC catalog, we considered the $\sim$ 6000 BX galaxies present in our narrow-band detected catalog. About 200 of them were characterized by spectroscopic redshift. We then chose the BXs characterized by $2.0<z_{spec}<2.2$ from the GOODS Mastercat \footnote{http://archive.eso.org/cms/eso-data/data-packages/goods-fors2-final-data-release-version-3.0}$^,$\footnote{http://archive.eso.org/cms/eso-data/data-packages/goods-vimos-spectroscopy-data-release-version-1.0/index} and MUSYC databases.  
We excluded galaxies that showed an IRAC 3.6 $\mu$m detected source within an annulus of 1$''$ internal and 4$''$ external radius from their position. As a result, this sample is characterized by 16 objects with comparable photometry to our LAE sample. Due to Ly$\alpha$ selection, most LAEs are faint in the continuum. The UV-bright LAE sub-sample is expected to overlap with BX galaxies in this property and so we investigate if it is reflected in the SED properties.

\section{SED FITTING METHODOLOGY}
\label{sec:model}

We used the S. Charlot \& G. Bruzual (2010, private communication, hereafter CB10) code to compute the stellar population synthesis models. The updated version of the original code includes the treatment of the thermally pulsating asymptotic giant branch (TP-AGB) stars (prescription of \citeauthor{MarigoGirardi2007} 2007). 
We used solar metallicity, for which the stellar population model was described as ``very good'' in the Bruzual \& Charlot (2003) documentation. In addition to that, the continuum model SED did not show significant differences in the shape when changing the metallicity value.
In order to be able to compare with previous works we adopted a Salpeter initial mass function (IMF, \citealt{Salpeter1955}),
distributing stars from 0.1 to 100 M$_{\odot}$, following $\Phi(M)\propto M^{-2.35}$. 
 However, a Chabrier IMF (\citealt{chabrier2003}), for example, being characterized by a flatter slope, might be more appropriate for deriving an absolute value of the stellar mass or SFR, while a Salpeter one can over-estimate these quantities by a factor 1.8 (\citealt{Papovich2010}). 
 
This choice of population synthesis model permitted us to add a star formation history (SFHs) parametrized by   
the $\tau$ parameter (\citealt{Maraston2010}, \citealt{Lee2010}),
\begin{equation}
{\rm SFR(t)}= \frac{N}{|\tau|} e^{-\frac{t}{\tau}}. 
\label{sfrh}
\end{equation} 
N was the normalization of the model which gave the observed stellar mass as a function of the star formation history.
Equation (\ref{sfrh}) allowed the star formation to be exponentially declining (positive value of $\tau$), exponentially increasing (negative value of $\tau$)
or constant ($\tau \rightarrow \infty$). 
As described in \citeauthor{Maraston2010} (2010), an exponentially increasing SFR could be a good representation of the star formation history of a starburst galaxy at $z\sim2$, the epoch of peak cosmic star formation density.\\ 
For $|\tau|>>$~t, the SFH approaches the constant star formation rate,\\ 
$lim_{\frac{t}{\tau} \to 0} \frac{N}{|\tau|} e^{-\frac{t}{\tau}} = \frac{N}{|\tau|}$.\\
For $0<\tau<<$~t, the model describes a single burst, with SFR~$\to0$ at late times, \\
$lim_{\frac{t}{\tau} \to \infty} \frac{N}{\tau} e^{-\frac{t}{\tau}} = 0$\\ 
For $\tau<-$t ($\tau>$~t), equation (\ref{sfrh}) can be approximated as a linear increase (decrease) in time, e.g., 
$lim_{\frac{t}{\tau} \to 0} \frac{N}{|\tau|} e^{\frac{-t}{\tau}} = \frac{N}{|\tau|}~(1-\frac{t}{\tau})$. \\
Therefore, this model is able to produce approximately constant, linearly increasing/decreasing, exponentially increasing/decreasing and single burst star formation histories. We will adopt the exponentially increasing, decreasing and constant star formation (CSF) as the three main scenarios of star formation histories. 

The principal parameters used for the SED fitting were the observed stellar mass, M$_*$, the extinction, E(B-V), the e-folding time, $\tau$, and the age since the beginning of star formation, age$_{\rm SF}$. These were the input parameters that went into the computation of each model's SED, to be compared with the data. We defined a four-dimensional grid of points, the ``parameter space'', by sampling several tens of values for each parameter. We used a logarithmic spacing for age$_{\rm SF}$, which varied between 0.005 Gyr and 3 Gyr, also for $\tau$, which varied between 0.005 Gyr and 4 Gyr, as well as for the observed stellar mass, whose range was $10^6-10^{12}$ M$_{\odot}$. The age of the Universe at $z \simeq 2.1$ is $\sim 3$ Gyr.  For E(B-V), we adopted a linear spacing between 0.0 and 0.6. Tab. \ref{tab:freeparam} summarizes the parameter grid. 

We took the luminosity densities (erg s$^{-1}$ {\AA}$^{-1}$ per unit of solar luminosity) output by CB10, applied the Calzetti dust extinction law 
and a correction for the intergalactic neutral Hydrogen absorption using the Madau law (\citealt{Madau:1995}). We then convolved the flux densities with the filter transmission curves (Fig. \ref{filters}). 

For each model (identified by a point in the grid) we computed a $\chi^2$ value as described below. 
To find the best fit model and compute confidence levels of the fit parameters, we studied how the probability density changed along the grid. This was done by computing,  \begin{equation}
\chi^2=\sum_{i=1}^{13}\frac{(f^{\rm model}_{\nu,i}[M_*, E(B-V), \tau, age_{\rm SF}]-f^{\rm obs}_{\nu, i})^2}{\sigma^2_i},
\end{equation}
where $i$ ran over the 13 observed bands, $f^{\rm model}_{\nu, i}$ and $f^{\rm obs}_{\nu, i}$ were respectively the theoretically predicted (for each model) and observed flux densities, and $\sigma_i$ was the uncertainty on the $i$-th data point.

Once the $\chi^2$ was computed for each combination of parameters in the grid, we identified all models that had $\chi^2$ within the 68\% confidence level from the expected $\chi^2$ for 9 degrees of freedom. We rescaled the delta-$\chi^2$ corresponding to the 68\% confidence level, based on the $\chi^2$ of the best fit. This corresponded to the best fit in a Frequentist approach. For each parameter, we reported the 68\% confidence level range as the parameter uncertainty. 

CB10 produces the galaxy continuum luminosity density. The spectral energy distribution of a galaxy can also be influenced by nebular continuum emission,
which can make a significant difference when the galaxy is younger than 5 Myr, and nebular emission lines, which can affect the results for ages up to $\sim$ 50 Myr.  
Hot stars in active star-formation regions provide a large number of ionizing photons. They excite neutral Hydrogen (and also Helium and Oxygen) producing nebular emission lines, but also a gas continuum luminosity that is proportional to the Lyman continuum ($\lambda_{rest-frame} < 912$ {\AA}) photons. 
The relevant nebular emission lines produced in a star forming galaxy at $z=2.1$ are the Ly$\alpha$ emission itself which enters the observed $U$ band (we are taking into account its contribution, by subtracting Ly$\alpha$ flux from the $U$ continuum as in Gu10), [OII] $\lambda 3727$ in the $J$ band, H$\beta$ and [OIII] $\lambda4959, \lambda5007$ in the $H$ band, H$\alpha$ in the $K$ band, and Pa$\alpha$ in the 5.7 $\mu$m band. 

The inclusion of the nebular emission is beyond the scope of this paper. Here we adopted a standard IMF (Salpeter, 1955), a reference metallicity, Z$_{\odot}$, and a pure continuum in the SED model to be able to explore a more complicated SFH. We will include a proper treatment of the nebular emission in a subsequent paper (\citealt{Acq2011}). 
In any case, the nebular continuum is wavelength dependent and affects mainly the rest-frame optical. As its effect falls within our error bars, we can ignore its contribution to the SED at this stage. 
In addition to the nebular continuum, at $z\simeq2.1$ the nebular emission lines also enter observed bands that are characterized by high uncertainties (see Fig. \ref{fullsamplebestfits}). Therefore, they are unlikely to affect the SED results we show in \S4. 
We checked that when including nebular continuum and emission lines in a constant star formation model, the resulting SED fitting parameters were consistent within the 68\% confidence level for the FULLsample, analyzed without nebular emission. In future analyses, better NIR photometry is needed to better constrain the age of LAEs. At that point the nebular emission inclusion would be important, because models with only stellar continuum can appear older than those with the proper nebular gas treatment.

Nebular emission lines can also come from Helium. HeII emission at $\lambda =1640$ {\AA} enters the observed $B$ band.
Since HeII $\lambda 1640$ emission is weaker than Ly$\alpha$ (it is expected to have EW of just few {\AA}, Schaerer \& Vacca 1998, Brinchmann et al. 2008) and Ly$\alpha$ is a small part of the $U$ band flux, we can be confident that the $B$ band flux is not significantly affected by HeII $\lambda 1640$. 
We searched for HeII line detection ($\lambda_{observed-frame}=5028$ {\AA} at $z=2.066$), estimating the flux density of our LAEs in the NB5015 image of the ECDF-S (Ciardullo et al. 2010 in prep). The stacked flux density of the entire sample revealed non-detection of HeII. 
The average NB5015 flux is -0.0075 $\mu$Jy with a scatter of 0.058 $\mu$Jy. 
This can be translated into a mean EW(HeII $\lambda 1640$) of $-1.7\pm1.0$ {\AA}, where the errorbar gives the standard deviation of the mean derived from the observed scatter. A small equivalent width may be expected for young starburst galaxies. EW(HeII $\lambda 1640$) would be larger than 5 {\AA} in the case of an IMF characterized by a higher number of massive stars than a normal one, such as in the case of a top-heavy IMF. 
This unusual IMF would also produce objects with EW(Ly$\alpha) > 240$ {\AA} (\citealt{MR:2002}), which we did not find in significant number in our sample. We neither found objects with EW(HeII $\lambda 1640)>5 ${\AA} at 95\% confidence level.
In the distribution of EW(HeII $\lambda 1640$) of our $z\simeq2.1$ LAEs there is an equal number of objects characterized by strong positive and negative EW(HeII $\lambda 1640$), implying that all of these objects could be affected by photometric errors rather than true strong HeII emissions. Even if only spectroscopy can confirm this, we do not see evidence either for weak HeII emission from the population or for a significant number of individual objects with strong HeII emission. As a result, the non-detection of HeII flux density argues against the hypothesis of a top heavy IMF in the LAE population.

We reported the input parameters of the SED, M$_*$ in M$_{\odot}$, E(B-V) in magnitudes, $\tau$, and age$_{\rm SF}$ in Gyr, as free parameters of the SED fit. We also computed, as derived parameters, the mean stellar population age, t$^*$, in Gyr, the instantaneous SFR, the averaged SFR over the last 100 Myr of galaxy evolution, $<$SFR$>_{100}$, the SFR$_{corr}$(UV), obtained from the dust-corrected rest-frame UV flux density, in M$_{\odot}$ yr$^{-1}$, and the age$_{\rm SF}$/$\tau$ ratio. 
The t$^*$ parameter was determined to be the average of the age of the stars in a period corresponding to the star formation phase, 
\begin{equation}
t^*={age}_{\rm SF} - \frac {\int_0^{{age}_{\rm SF}} t \times  {\rm SFR}(t) dt} {\int_0^{{age}_{\rm SF}} {\rm SFR}(t) dt}
\label{tstar}
\end{equation}
The instantaneous SFR was calculated using Equation \ref{sfrh}, while the averaged SFR was defined as
\begin{equation}
<SFR>_{100}=\frac{\int_{\mathrm{100 Myrs}} SFR(t) dt}{\mathrm{100 Myr}}.
\label{aSFR}
\end{equation} 
We chose 100 Myr, because this was the minimum SF time for which the SFR$_{corr}$(UV) estimator was robust (\citealt{Kennicutt:1998}). 
We calculated the latter from the rest-frame UV (1500-2800 {\AA}, observed $R$ band) luminosity density (Equation 1 in Kennicutt, 1998) and assuming the E(B-V) range allowed by the SED fit for the dust-correction (Gu10). Kennicutt's formula assumed a constant star formation rate and that the galaxy was at least 100 Myr old.  For a 10 Myr old galaxy, the inferred SFR was estimated to be roughly half the real one. Therefore, for a CSF model, $<$SFR$>_{100}$ was directly comparable with SFR$_{corr}$(UV).
However if the galaxy was younger than 100 Myr, $<$SFR$>_{100}$ and SFR$_{corr}$(UV) were under-estimations of the on-going SFR.
Also, for an exponentially decreasing SFH, we would already expect that the instantaneous SFR is lower than $<$SFR$>_{100}$. Instead, assuming an exponentially increasing SFH, we would expect an instantaneous SFR larger than $<$SFR$>_{100}$.

The age$_{\rm SF}$/$\tau$ derived parameter was determined to indicate the number of e-foldings that the SFH has undergone. In the case of exponentially decreasing SFR, a large value of age$_{\rm SF}$/$\tau$ implied passive evolution with a negligible ongoing SF and that the majority of the stars in the galaxy were old (\citealt{Grazian2007}).
A large value of age$_{\rm SF}$/$\tau$ in the exponentially increasing SFR case meant the age$_{\rm SF}$ was a poorly constrained parameter since the implied old stellar population was a very small fraction of the mass.
The total ``stellar'' mass of a galaxy, which includes the mass of stellar remnants and outflows but does not account for return of outflows into future generations of stars, can be calculated analytically from the integral of SFR(t), Mgal(age$_{\rm SF}$)=$\int_{0}^{age_{\rm SF}}$ SFR(t) dt.   
This turned out to be 10-20\% higher than the observed stellar mass, M$_*$.

\subsection{Median statistics in SED fitting analysis}
\label{sec:median}

As most individual LAEs in our sample had insufficient S/N to allow meaningful SED parameter estimation, the only way to learn about the majority of the sample was to analyze their stacked SED.
A few galaxies with sufficiently bright continuum could be fitted individually, but, doing this would bias the results in favor of the properties of objects with bright continuum. The stacking approach was also adopted by Lai et al. (2008) and Gawiser et al. (2007) to fit the sample of ECDF-S LAEs at $z\simeq3.1$.
Outliers in the LAE distribution and other contaminants are characterized by observed SEDs different from the $typical$ one. As the mean stack of the individual SEDs would be sensitive to outliers, we stacked the measured LAE fluxes (including formal non-detections and negative fluxes) by computing the median flux in each band.
The analysis of the full sample reveals the property of the $typical$ LAE galaxy; to investigate $heterogeneity$ in physical properties that might be erased through our stack procedure we stacked sub-samples based on photometric properties (\S \ref{sec:subsamples}).

We created 200 model galaxy SEDs with M$_{*}$, E(B-V), $\tau$, and age$_{\rm SF}$ parameters normally distributed around M$_* = 2.9 \times 10^9$ M$_{\odot}$, E(B-V) = 0.21, $\tau$ = 0.1 Gyr, and age$_{\rm SF}$ = 0.053 Gyr. These represent typical $z\simeq2.1$ LAEs in our observed sample. We added 40 (15\%) simulated $z\simeq2.1$ galaxies with very different properties with respect to the typical ones, and 20 (8\%) low-redshift interlopers.
We generated a median stack of SED fluxes and associated error bars to the fluxes in the same way used for the real data, including the sample variance error, evaluated via the bootstrapping technique, and the photometric errors. The results of the SED fit were M$_* = 2.82 [1.78 - 3.16] \times 10^9$ M$_{\odot}$; E(B-V) = 0.22[0.20 - 0.28];  ; $\tau$ = 0.087[(-2.952)- (2.759)] Gyr; age$_{\rm SF}$ = 0.057 [0.023 - 0.143] Gyr.  

We also generated 200 model galaxy SEDs starting with input parameters uniformly distributed in the ranges $8.3\leq$ log(M$_*$/M$_{\odot}) \leq 9.9$, $0.00\leq$ E(B-V) $\leq 0.29$, $-2.3\leq$ log($\tau) \leq 0.6$, $-2.05\leq$ log(age$_{\rm SF}) \leq -0.64$. The median input parameters were log(M$_*$/M$_\odot$) = 9.1, E(B-V) = 0.16, log($\tau$) = -0.87, log(age$_{\rm SF}$) = -1.34. Then we included outliers as objects characterized by much lower and much higher stellar mass at $z\simeq2.1$ and located at lower redshift. We stacked the corresponding fluxes calculating their median and fitted the resulting model SED. We recovered the four parameter ranges allowed within the 68\% confidence level, log(M$_*$/M$_{\odot}$) =  [8.95 - 9.45]; E(B-V) = [0.08 - 0.25];  log($\tau$) = [(-2.3)- (0.6)]; log(age$_{\rm SF}$) = [(-1.9) - (-0.79)]. They correspond to the 68\% of the input ranges. The median stacked SEDs are able to reproduce the typical properties of a sample even in the presence of outliers and low-redshift interlopers.

In addition to that, to test the reliability of our code results, we ran it on Mock catalogs, generated from CB10 models fixing particular values of the free parameters. In all the tested cases ($\tau>0$, $\tau<0$, age$_{\rm SF}>|\tau|$, and age$_{\rm SF}<|\tau|$) we could recover the original parameters of the models within the 68\% confidence level.

\section{RESULTS}
\label{sec:results}

We fit the observed stacked SEDs of the full sample of $z\simeq2.1$ LAEs and those of the sub-samples enumerated in \S \ref{sec:subsamples}. 
 
\subsection{FULL sample}
\label{sec:fullsample}

In Fig. \ref{fullsamplebestfits} we show the observed SED of the full sample of $z\simeq2.1$ LAEs. The best fit model SED is obtained for an exponentially increasing star formation model. It is characterized by $\chi^2=12.3$ for 13 data points for the combination of parameters labelled in red in the figure.
The exponentially decreasing and constant star formation histories (SFH) produce almost equally good fits. Four models at the edge of the 68\% confidence level ($\chi^2\simeq$21.94) are also shown. They are chosen to be characterized by the highest allowed M$_*$ ($\sim$10$^9$ M$_{\odot}$) with E(B-V)=0.07 or E(B-V)=0.1 and lowest allowed M$_*$ (2.82 $\times$ 10$^8$ M$_{\odot}$) values with E(B-V)=0.16 or E(B-V)=0.17. The other models allowed at the 68\% confidence level have SED shapes within these extremes.

Tab. \ref{tab:solofull} shows the best fit and 68\% confidence level results for the entire parameter space and for the three star formation histories separately. In addition to the free parameters, it also reports the instantaneous SFR and $<$SFR$>_{100}$. 
For the best fit model, we obtain an instantaneous SFR=35[0.003-170] M$_{\odot}$ yr$^{-1}$ and a $<$SFR$>_{100}$=4[2-30] M$_{\odot}$ yr$^{-1}$.  

M$_*$ and E(B-V) are quite well constrained while there is a wider uncertainty in age$_{\rm SF}$ and essentially all values of $\tau$ are allowed at 68\% confidence. This is more easily seen in Fig. \ref{6panel}, where we show the contours of the allowed parameter space regions for the three main SFHs.
In the figure we can see the trends of E(B-V) versus age$_{\rm SF}$ (top left panel) and versus $\tau$ (top right panel).
Reddening and age$_{\rm SF}$ are mildly 
degenerate; the shape of the SED can be similar if the reddening is higher or if the spectrum is older. 
The middle right panel shows M$_*$ versus E(B-V), while the bottom right panel shows 
$\tau$ versus age$_{\rm SF}$.
For the exponentially decreasing SFH, the best fit age$_{\rm SF}$ and $\tau$ are both close to 10 Myr. However, the allowed parameter space is uniformly covered in the age$_{\rm SF}$ and $\tau$ directions. For the exponentially increasing SFH, the best fit age$_{\rm SF}$ and $|\tau|$ are both close to 20 Myr. High values of $|\tau|$ together with high age$_{\rm SF}$ values are disfavored 
at the 68\% confidence level. Models with age$_{\rm SF}$ up to 1 Gyr are allowed for SFHs corresponding to $|\tau|$ below 100 Myr. Models with age$_{\rm SF}$ up to 3 Gyr are allowed for $|\tau|$ around 100 Myr.

\subsubsection{Comparison with $z\simeq3.1$ LAEs}

In Tab. \ref{tab:ALLcomp} we compare the SED fitting parameters with others published for LAEs at redshifts 2 and 3.
Previous work by \citet{Nilsson:2007}, hereafter N07 and \citet{Ono2010}, hereafter O10 assumed a constant SFR history, so we focus on that model for comparison. 
In the case of N07, we report the average behavior of the $z\simeq3.1$ fitted galaxies; in the case of O10, we report the parameter range of their $K$-detected galaxies and the results from the stack of the $K$-undetected sources. 
\citet{Gawiser:2007}, hereafter Ga07, fit the stacked SED of the sources with non-detection at 2$\sigma$ in the 3.6 $\mu$m IRAC band image. They used a two-population model with each population having an exponentially decreasing SFH.
Nilsson et al. (2009) reported that the preliminary SED fitting of their $z\simeq2.3$ LAEs shows they are older and dustier than their $z\simeq3.1$ sample, noting a wider range of possible fitting models. 
Such a result was recently expanded upon in \citeauthor{Nilsson2010} (2011, hereafter N10). 
They estimate the age of the old and of the young stellar population and the mass fraction in the young component. As for their entire sample the mass fraction in the young component is 56\%, we reported the young age in the table. 
While a direct comparison of ages and star formation histories with our work is not possible since they use two Simple Stellar Population (single-burst) models and two stellar populations, they confirmed the earlier finding of a higher amount of dust at lower redshift, and report high stellar masses of [5-10] $\times$10$^9$ M$_{\odot}$. We see that our LAEs at $z\simeq2.1$ are dustier than those at $z\simeq3.1$ and than the $K$-detected galaxies from O10, while the observed stellar masses are consistent with N07 and Ga07. They are significantly higher than the masses of the sample of $K$-undetected sources from O10 and lower than the range of values obtained by the same authors for the $K$-detected sources. The age$_{\rm SF}$ of our sample is consistent with the results for $K$-undetected sources from Ono et al. (2010), but a wide range of values is allowed. 
 Our estimated SFR is higher due to the higher inferred dust extinction and lies in the range of values found by O10 for their $K$-detected sources. We will compare more directly the Lai et al. (2008) results with our IRAC-bright and IRAC-faint samples in  
\S \ref{sec:iracbrightfaintCOMP}.

\subsection{IRAC bright and faint sub-samples}  
\label{sec:iracbrightfaint}

The first sub-samples we fit are those selected based upon the IRAC 3.6 $\mu$m flux (Tab. \ref{tab:subsampledefinition}). 
In Fig. \ref {SEDpanel} (upper left panel) we present the observed and best fit model SED for the IRAC-bright and IRAC-faint galaxies. 
The IRAC-faint sub-sample is characterized by a flat rest-frame UV-slope and thus by a low dust amount; the observed (N)IR bands present a higher flux for the IRAC-bright sources,  implying an order of magnitude higher observed stellar mass. This is consistent with Fig. \ref{contsubsample}, that shows stellar mass versus reddening contours of the sub-samples at 68\% confidence level. These contours do not overlap for IRAC-bright and IRAC-faint galaxies. 
Therefore, the IRAC-faint galaxies appear to be significantly less massive and less dusty than the IRAC-bright. 
In Fig. \ref{contageE} the contours indicate the E(B-V) versus age$_{\rm SF}$ allowed regions for the sub-samples.  
While the E(B-V) 68\% confidence level regions show the IRAC-bright typical object is significantly dustier than IRAC-faint, the age$_{\rm SF}$ contours overlap.
 
The IRAC-faint $\tau$ versus age$_{\rm SF}$ contours (Fig. \ref{conttauage}) occupy almost the entire parameter space. The IRAC-bright contours show a different behavior in the case of exponentially decreasing or increasing SFHs. When fitting the data with an exponentially decreasing model, age$_{\rm SF}$ is better constrained ($<20$ Myr), while with an exponentially increasing model higher age$_{\rm SF}$ values are allowed, but just for $|\tau|$ shorter than 100 Myr. 

Tab. \ref{tab:bestfitparameter} presents results for free and derived parameters with 68\% confidence levels for the sub-samples. We can see that 
the IRAC-bright galaxies' SED is consistent with a $<160$ Myr old population, with an $<$SFR$>_{100}$ of [10-120] M$_{\odot}$ yr$^{-1}$ and an 
instantaneous SFR of [40-1310] M$_{\odot}$ yr$^{-1}$. 
The IRAC-faint sub-sample is characterized by a stellar population of  age$_{\rm SF}\leq1.5$ Gyr, with a lower $<$SFR$>_{100}$ of [1-15] M$_{\odot}$ yr$^{-1}$.   

\subsubsection{Comparison with IRAC detected and undetected $z\simeq3.1$ LAEs}
\label{sec:iracbrightfaintCOMP}

In Tab. \ref{tab:ALLcomp} we also compare the IRAC-faint and IRAC-bright sub-sample results with the Lai et al. (2008) L08det and L08und results. 
Lai et al. used the Bruzual \& Charlot (2003) version of the stellar population synthesis model in their analysis and assumed a CSF. Our sub-sample luminosity threshold matches theirs, allowing for a fair test of redshift evolution.
We can see that the E(B-V) estimation obtained for our sample implies a higher dust amount at $z\simeq2.1$ than at $z\simeq3.1$. 
The observed stellar masses are consistent between the $z\simeq2.1$ IRAC-faint and $z\simeq3.1$ L08und sub-samples, but are significantly smaller for the IRAC-bright sub-sample than they were for the L08det. Lai et al. (2008) found that L08det galaxies were older and more massive than 
L08und ones, but that both categories of galaxies were dust-free. Here we find that IRAC-bright are dustier and more massive than IRAC-faint, with IRAC-bright requiring non-zero dust.

\subsection{Red LAE sub-sample}
\label{sec:redblue}
 
In the upper right panel of Fig. \ref{SEDpanel} we show the observed and best fit model SEDs for the sub-sample of red-LAEs. 
They are significantly more massive and dustier than the typical galaxies, as is also seen in Fig. \ref{contsubsample}. 
It is interesting to note that in spite of being dusty and massive, the stellar population of the red sub-sample can be as young as a few tens of Myr, implying that the population dominating the SED is still a young one, with a very high instantaneous SFR ([1-2880] M$_{\odot}$ yr$^{-1}$, within the entire parameter space). 
In Fig. \ref{contageE} we can see that the E(B-V) allowed ranges are significantly different for the red with respect to typical LAEs, while the age$_{\rm SF}$ ranges overlap. Fig. \ref{conttauage} shows that only a few locations in the $\tau$ versus age$_{\rm SF}$ parameter space are disfavored for the red-LAEs. However, exponentially increasing models with age$_{\rm SF}$/$|\tau|$ ratio close to 1 are ruled out at the 68\% confidence level, for age$_{\rm SF}>0.1$ Gyr.
Their SF phase could be as long as 2 Gyr (Tab. \ref{tab:bestfitparameter}).

\subsection{Rest-frame UV bright and faint sub-samples}
\label{sec:UVbrightUVfaint}
 
The rest-frame UV-bright and faint sub-samples (Tab. \ref{tab:subsampledefinition}) are analyzed here. In the lower left panel of Fig. \ref{SEDpanel} we show  
that the UV-faint galaxies can be characterized by a low-dust SED, with a quite flat rest-frame UV. 
The contour plots in Fig. \ref{contsubsample} and Fig. \ref{contageE} also show that the UV-bright typical galaxy is more massive and dustier than the UV-faint. 

The UV-bright objects are characterized by $R<25.5$. As with the IRAC-bright sub-sample, brighter continuum enables better age determination. In fact for the typical UV-bright LAE the allowed region of the age$_{\rm SF}$ is as narrow as [0.007-0.404] Gyr within the entire parameter space (Tab. \ref{tab:bestfitparameter}) and [0.007-0.064] Gyr in the case of the exponentially decreasing model (Tab. \ref{tab:last}). However, for the UV-faint SED basically the entire age$_{\rm SF}$ range is allowed.
 In Fig. \ref{conttauage} the $|\tau|$ versus age$_{\rm SF}$ parameter space seems to follow the pattern seen for the IRAC-bright sub-sample. The exponentially increasing models with $|\tau|>100$ Myr are ruled out for age$_{\rm SF}>20$ Myr.

The instantaneous and averaged SFRs can be much higher for the UV-bright than the UV-faint sub-sample. 
This is expected because the SFR is proportional to the rest-frame UV flux, traced by the observed $R$ band, multiplied by a correction for dust extinction.

\subsection{High-EW and low-EW sub-samples}
\label{sec:ewhighlow}

The high-EW and low-EW sub-sample (Tab. \ref{tab:subsampledefinition}) results are presented here. 
The lower right panel of Fig. \ref{SEDpanel} shows their observed and best fit model SEDs.

In Fig. \ref{contsubsample}, \ref{contageE} and \ref{conttauage} we can see that the 68\% confidence level contours of these sub-samples overlap.
However, the typical low-EW galaxy shows slightly higher stellar masses (10$^{8.5}$-10$^{9.2}$ M$_{\odot}$) than the high-EW (10$^{7.5}$-10$^{8.9}$ M$_{\odot}$). 
The high-EW LAEs appear to be the least massive objects (Fig. \ref{contsubsample}).
This could be caused by these galaxies experiencing one of their first phases of star formation and thus not yet having formed as much stellar mass as typical LAEs.
The E(B-V) allowed ranges seem to be consistent for both sub-samples (see also Tab. \ref{tab:bestfitparameter} and \ref{tab:last}), revealing that high-EW LAEs can be characterized by a moderate amount of dust, even if models implying the absence of dust also produce acceptable fits.

$<$SFR$>_{100}$ can be five times higher for the low-EW than for the high-EW sub-sample. The stellar population can be as old as 1.5 Gy for the high-EW, but appears younger than 900 Myr for the low-EW galaxy. 

We also examine the dependence upon Ly$\alpha$ luminosity, splitting the galaxies with L$_{Ly\alpha}>10^{42.1}$ erg sec$^{-1}$ in half. We do not find significant differences in the allowed ranges of the SED parameters. 
We can compare our reliable estimates of M$_{*}$ and E(B-V) for the L$_{Ly\alpha}>10^{42.1}$ erg sec$^{-1}$ objects with the recent results by N10 at $z\simeq2.3$ (which reached a factor of 1.8 less deep Ly$\alpha$ luminosity limit). Looking at the entire parameter space we find agreement with their results for dust reddening, but our SED analysis implies a stellar mass roughly one order of magnitude lower.

\subsection{BX star forming galaxies}
\label{sec:bx}

In Fig. \ref{BX} we show the observed and best fit SEDs obtained from stacking a sample of ECDF-S star forming galaxies at $z\simeq2.1$ (\S \ref{sec:BX}).
Within the entire parameter space, the SED best fit is obtained for an exponentially increasing SFH (brown solid line). The best fit and allowed parameter ranges are 
log(M$_*$/M$_{\odot}$)=10.0[9.5-10.1], E(B-V)=0.1[0.0-0.4], and age$_{\rm SF}$=0.09[0.01-0.20] Gyr. Our data for BX cannot discriminate between different SFHs. 
The dotted cyan and solid blue lines overlap and represent the exponentially decreasing and CSF best fit models.  
The magenta dashed line shows the Erb et al. (2006) SED, assuming the median best fit parameters of their sample to build the model spectral energy distribution. This is a very good fit for the rest-frame UV part of the spectrum, but corresponds to a much higher IR flux density, producing a $\chi^2 \sim80$. 

As a comparison we show the best-fit model SEDs (green solid line) for the UV-bright LAEs, selected to have the same observed $R$ band magnitude limit as the BXs. They present a much flatter rest-frame UV continuum, caused by a much lower reddening and/or also to a younger stellar population. Our limit in the NIR does not allow us to constrain the age$_{\rm SF}$ parameter or to discriminate between the two possibilities.
The typical UV-bright LAE (log(M$_*$/M$_{\odot}$)=9.1[8.9-9.4] from Tab. \ref{tab:bestfitparameter}) appears less massive than either our BX or Erb et al. (log(M$_*$/M$_{\odot}$)=10.3[10.2-10.4]) star forming galaxies.

In Tab. \ref{tab:ALLcomp} we present the best fit and allowed parameter ranges for a CSF model to compare with $z\simeq2.1$ and $z\simeq3.1$ LAE literature.
Although the small BX sample is characterized by large sample variance, the typical BX galaxy appears dustier and more massive than our LAEs at $z\simeq2.1$ and previously studied LAEs at $z\sim3$. Their higher stellar mass implies a more evolved stage than LAEs, perhaps as a consequence of previous phases of star formation. LAEs may occupy the low mass tail of the BX mass function.

\section{DISCUSSION}
\label{sec:discussion}

\subsection{SED parameter constraints}
\label{sec:constraints}

Due to the smaller number of objects in the sub-samples, their observed SEDs are characterized by higher uncertainties than the FULLsample.  
Although each sub-sample contains a significant fraction of objects not included in any other sub-sample, Tab. \ref{tab:overlap} shows sufficient overlap that similarities in the SED properties could be expected for the typical UV-faint(bright) and IRAC-faint(bright) sub-samples. 
Since we are considering the median stacked SEDs, it is worthwhile to study the SED fitting parameters of both of them. 

We presented the SED stacking and fitting results in Fig. \ref{fullsamplebestfits} for the FULLsample and in Fig. \ref{SEDpanel} for the four pairs of sub-samples, with the FULLsample best fit model SED drawn for comparison.
We do not possess enough signal-to-noise to fit all, or even most, galaxies individually.  
However, our sub-sample analysis allow us to investigate the $heterogeneity$ of our sample, i.e., whether there are differences between the typical sub-sample member and the typical $z\simeq2.1$ LAE. In Fig. \ref{SEDpanel} the observed SEDs already seemed significantly different between pairs of sub-samples.
In all sub-sample SEDs, the rest-frame UV fluxes are characterized by smaller error bars, while the N(IR) bands have larger uncertainties (Tab. \ref {tab:limitingMAG}). For this reason the physical properties we can obtain from the rest-frame UV spectral range are well constrained. 
The solid lines in the figure corresponded to the SEDs built using the best fit M$_{*}$, E(B-V), $\tau$, age$_{\rm SF}$ parameters listed in Tab. \ref{tab:bestfitparameter}.  
The choice of parameter pairs in Fig. \ref{contsubsample}, \ref{contageE}, and \ref{conttauage} was made to show the contours of two well constrained, one constrained and one non constrained, and two non constrained parameters, respectively.

\subsubsection{Stellar mass, M$_{*}$, and reddening, E(B-V)}
\label{sec:ME}
 
M$_{*}$ and E(B-V) are well constrained. In general, we do not see a strong correlation between E(B-V) and stellar mass. 
Looking at horizontal pairs of panels in Fig.  \ref{contsubsample}, we conclude that the typical IRAC-bright LAE is significantly more massive and dustier than the IRAC-faint, that the UV-bright LAEs are slightly more massive and significantly dustier than the UV-faint, and that stellar mass and reddening contours overlap for the typical high-EW and low-EW galaxies. Looking at the figure along the vertical direction, it appears that the red-LAEs assume the highest stellar mass and reddening values, while the high-EW galaxies have the lowest stellar mass and dust amount among all the sub-samples.

\subsubsection{Star formation age and e-folding time}
\label{sec:age}

The age since star formation began, age$_{\rm SF}$, and the e-folding time, $\tau$, are very difficult to constrain with our data.
One of the reasons why the age$_{\rm SF}$ parameter is difficult to constrain is the high uncertainty seen in the observed-frame NIR bands, where the Balmer break falls. The age of a young stellar population that dominates the rest-frame UV continuum (and, presumably, the Ly$\alpha$ photons) is, thus, represented by the mean stellar age parameter, t$^*$, with the overall age since the beginning of star formation being very difficult to determine due to the low luminosity of evolved stellar populations.  
When considering stacks of observed SEDs with higher S/N in the continuum, age$_{\rm SF}$ appears better determined. This is, for example, the case of the stack of BX continuum selected star forming galaxies (\S \ref{sec:bx}, age$_{\rm SF}=$~[10-200] Myr).
In general we see that age$_{\rm SF}$ appears better constrained in the exponentially decreasing SFH and in particular in the CSF case, due to the loss of one degree of freedom via the constraint of M$_*=$age $\times$ SFR. When assuming an exponentially increasing SFH, the large amount of recently formed stars, which dominate the model SED, make it nearly impossible to determine the age$_{\rm SF}$ in the galaxy.

In many cases, similarly good best fits can be obtained from $\tau>0$, $\tau<0$, or constant SFR models and slightly different combinations of M$_{*}$, E(B-V) and age$_{\rm SF}$. This implies that the S/N in the data does not allow us to single out one of them.

Although our photometry does not allow us to constrain age$_{\rm SF}$ and $\tau$, Fig. \ref{conttauage} shows that there are regions of the parameter space which are disfavored at the 68\% confidence level.  
For the IRAC-bright and UV-bright sub-samples, models older than $>20$ Myr and with a $|\tau|$ parameter higher than 100 Myr are ruled out. 
As a consequence, t* is required to be $\leq160$ Myr for these sub-samples.

Previous work fitting SEDs of star forming galaxies with an exponentially declining SFR also found the star formation e-folding parameter difficult to constrain (e.g., \citealt{Papovich2001}). 
\citeauthor{Pentericci:2007} (2007) fitted a set of $z\sim3$ LBGs, some of which show Ly$\alpha$ in emission. The age$_{\rm SF}$ and $\tau$ parameters they found were allowed almost in their entire range, but they were able to constrain the age$_{\rm SF}$ over $\tau$ ratio which they called the evolutionary stage of star formation. An age$_{\rm SF}/\tau > 4$ was characteristic of an evolved galaxy, assuming a decreasing SFR model (\citealt{Grazian2007}). 
In the same hypothesis, we find a median value of age$_{\rm SF}$/$\tau$ = 0.1 for the FULL sample and a similarly low value for all the sub-samples. 
However, the large uncertainties on age$_{\rm SF}$ and $\tau$ reflect in a wide range of allowed age$_{\rm SF}$/$\tau$ ratios for all SFHs, as shown in Tab. \ref{tab:bestfitparameter} and \ref{tab:last}.
\citeauthor{Erb:2006} (2006) used the exponentially decreasing SFR model to find the best fit SED of individual star forming galaxies at $z\sim2$. As they found that the $\tau$ parameter space was basically allowed in its entirety, they chose the star formation history that allowed them to recover SFR values in agreement with H$\alpha$ SFR estimations. 
We are now motivated to estimate the SFR averaged over the last 100 Myr (Equation \ref{aSFR}) and compare it with SFR$_{corr}$(UV) and SFR. We will present it and comment about the (dis)agreement in the next section.

\subsubsection{Star formation rate estimations}
\label{sec:sfr}

The instantaneous SFR is very sensitive to values of both age$_{\rm SF}$ and $\tau$. We therefore expect high uncertainties in its estimation as reported in Tab. \ref{tab:solofull}  
- \ref{tab:last}. 
In the majority of the cases we report its value for completeness, even if the large error bars do not make it a useful quantity to compare with the literature. 
Assuming a CSF, we expect the instantaneous SFR, the averaged SFR over the lifetime of the galaxy, and the SFR$_{corr}$ to agree.  
However, if a galaxy is less than 100 Myr old, $<$SFR$>_{100}$ would be lower than the SFR averaged over the lifetime of the galaxy. The latter is expected to be smaller than the instantaneous SFR by a factor roughly corresponding to 100 Myr/age$_{\rm SF}$. 
This behavior can be observed for instance in the FULLsample for each of the three SFHs (Tab. \ref{tab:solofull}), where the instantaneous SFR is higher than $<$SFR$>_{100}$ 
within the 68\% confidence level range.
On the other hand, SFR$_{corr}$(UV) = 15[6-30] M$_{\odot}$ yr$^{-1}$ is roughly consistent with $<$SFR$>_{100}$ = 4[2-9] M$_{\odot}$ yr$^{-1}$
 
A similar behavior is observed in the sub-samples. For the ones with younger SEDs (IRAC bright, red-LAEs, UV-bright), $<$SFR$>_{100}$ and SFR$_{corr}$(UV) imply lower values of star formation rate than the instantaneous SFR. It is interesting to note that for red-LAEs, the inferred values of instantaneous SFRs are as high as thousands M$_{\odot}$ yr$^{-1}$. For the sub-samples with older SEDs (IRAC-faint, UV-faint, high-EW, where the age$_{\rm SF}$ can reach several hundreds of Myr), the three SFR indicators are consistent. 
 
In \S \ref{sec:Xray} we will compare the SFR estimations of the FULLsample with that inferred by X-ray emission.

\subsection{LAE colors}
\label{sec:lae}

The red ($B-R\geq0.5$) sub-sample contains only 15\% of our sample LAEs, indicating that the majority of our $z\simeq2.1$ LAEs have blue rest-frame colors. The red-LAE sub-sample has the highest IR flux of any sub-sample as we can see in Fig. \ref{SEDpanel}. 
A more global description of the UV-through-IR SED is offered by the $R$- [3.6] (rest-frame NUV-$J$) color which is sensitive to the UV slope and the Balmer/4000{\AA} break, so it traces a combination of dust reddening and age. The 3.6 $\mu$m magnitude is a tracer of the stellar mass. 
We plot the $R$- [3.6] color versus the 3.6 $\mu$m band magnitude to study the color$-$stellar mass trend of the $z\simeq2.1$ LAEs and their sub-samples in Fig. \ref{R36R}a. 
It appears to be a linear relation between the mass and this color, with the IRAC-faint, UV-faint, and high-EW also being the bluest and least massive sub-samples. The red-LAEs appear at the upper left corner of the figure being a special sub-sample of extremely massive and dusty galaxies. They can be even dustier than spectroscopically confirmed $z\simeq2.1$ BX galaxies although their stellar masses can be similar. The BX are brighter in $R$ due to photometric and spectroscopic selection, so they lie outside the linear mass-color relation. 
L08 found a continuum of properties in this color-magnitude plot between $z\sim3$ LAEs and LBGs, observing that $z\simeq3.1$ LAEs were at the faint, blue end. 
We can say that $z\simeq2.1$ LAEs have a wide range of masses and there is a continuum of properties from the redder and more massive to the bluer and less massive sub-samples. Also, the datum corresponding to L08und sub-sample is located outside the linear mass-color relation, showing the less dusty and/or younger nature of the typical $z\simeq3.1$ LAEs than those at $z\simeq2.1$.

In Fig. \ref{R36R}b we show the $R$- [3.6]  color versus the $R$ magnitude, which is roughly proportional to the SFR derived from the rest-frame UV continuum assuming no dust extinction. The intrinsic SFRs calculated at $z\simeq2.1$ are much higher than at $z\simeq3.1$ but $z\simeq2.1$ LAEs are dustier. In this figure the dustiest red-LAEs are as faint as the IRAC-bright and UV-bright sub-samples in the rest-frame UV continuum.
Therefore, we see a less clear linear mass-SFR relation. The BX galaxies are the brightest objects as, by definition, they are selected to have $R<25.5$.
From this figure we can also see that the $R$ band sensitive SFR is comparable in the L08det, L08und, and our $z\simeq2.1$ LAE sub-samples, but the UV-bright, IRAC-bright, and red-LAEs are characterized by higher SFR values.

\subsection{Dust properties in LAEs}
\label{sec:dust}

In this sub-section, we investigate the dependence of the dust on photometric properties, such as L$_{Ly\alpha}$, EW(Ly$\alpha$).  

We find that wherever the dust amount is significant, the Ly$\alpha$ EW tends to be low. 
In fact the median(EW(Ly$\alpha))=30$ {\AA} for the typical IRAC-bright (E(B-V)=[0.3-0.5]), while it is 70 {\AA} for the typical IRAC-faint ((E(B-V)=[0.0-0.3])) galaxy. 
Galaxies brighter in the continuum, with M$_*>10^9$ M$_{\odot}$ and characterized by a significant amount of dust, show low equivalent width Ly$\alpha$ lines.
This result seems to be in disagreement with the hypothesis of enhancement of EW(Ly$\alpha$) due to the presence of a clumpy interstellar medium (\citealt{neufeld:1991}, \citealt{Fin:2008}). In this scenario, dust grains enclosed inside clumps of HI would not be able to absorb Ly$\alpha$ photons, as the HI density gradient would scatter them away from clouds, favoring their escape from the galaxy.

We relate the dust amount to the rest-frame UV slope and we also compare these quantities with the equivalent width and Ly$\alpha$ luminosity.
The slope of the rest-frame UV spectrum is proportional to the reddening of the stellar continuum measured by the E(B-V) parameter\footnote{Note that our quantity E(B-V) is referred to as E$_s$(B-V) by Calzetti et al. (2000), while the color excess derived from the nebular gas emission lines is defined as E$_g$(B-V).}, we approximate the flux density as a power law (f$_{\lambda}\propto\lambda^{\beta_{UV}}$) and estimate its index, $\beta_{UV}$. 
We fit the rest-frame UV spectrum from $B$ to $I$ to obtain $\beta_{UV}$ and its uncertainty. 
We use Calzetti's assumption that A(1600 {\AA}) = 4.39 $\times$ E$_g$(B-V), E(B-V) = (0.44$\pm$0.03) $\times$ E$_g$(B-V), and the \citealt{Meurer1999} relation A(1600 {\AA})~$= 4.43+1.99\times\beta_{UV}$. Therefore, high-E(B-V) objects are characterized by high, even positive, $\beta_{UV}$ values. E(B-V)$=$0 corresponds to f$_{\lambda}\propto\lambda^{-2.2}$, nearly flat in f$_{\nu}$.

In Fig. \ref{beta} we show the $\beta_{UV}$ index as a function of L$_{Ly\alpha}$ and EW$_{rest-frame}$.
From the left panel we can see that for higher values of Ly$\alpha$ luminosity (L$_{Ly\alpha}>$~ 10$^{42.35}$ erg sec$^{-1}$) $\beta_{UV}$ tends to -1.9 with a scatter of 0.2. 
This means that high-luminosity LAEs at $z\simeq2.1$ present a flat rest-frame UV spectrum. 
In the low-luminosity bin $\beta_{UV}$=-1.5 and the scatter is much larger ($\sigma_{\beta}$ = 1.1). This is due to the low L$_{Ly\alpha}$ objects that can be characterized by $\beta_{UV}$ up to 3. 
Tracing a horizontal line at $\beta_{UV}$ = -1.9 and a vertical line at log(L$_{Ly\alpha}$) = 42.35, 62\% of the less luminous and 45\% of the more luminous objects have $\beta_{UV}>$~ -1.9.
The right panel shows that the highest-$\beta_{UV}$ sources are red-LAEs, with EW$_{rest-frame}<50$ {\AA} and steeper spectra, mean($\beta_{UV}) = 0.17 \pm 1.15$.
Also, the UV-bright LAEs present  EW$_{rest-frame}<100$ {\AA} and $\beta_{UV} = -1.2 \pm 0.95$, while the objects brightest in Ly$\alpha$ luminosity are characterized by a flatter spectrum ($\beta_{UV}= -1.9 \pm 0.3$) with small scatter in $\beta_{UV}$ and EW$_{rest-frame}$ up to 200 {\AA}. 
In the same figure we also show the data from Nilsson et al. (2009) at $z\simeq2.3$. In their sample, LAEs with the highest EW values are also characterized by flat rest-frame UV spectra and high Ly$\alpha$ luminosities. Steeper spectra are seen for $z\simeq2.3$ EW$_{rest-frame}<40$ {\AA} objects, which have lower Ly$\alpha$ luminosity.

 In Fig. \ref{RE} we plot the $R$ magnitude versus the stellar continuum dust reddening, E(B-V), estimated from the ratio between the observed SFR(UV) and SFR(Ly$\alpha$) from Gu10 for the individual LAEs of our $z\simeq2.1$ sample. The observed SFR(UV) and SFR(Ly$\alpha$) are related to the intrinsic SFR by assuming the dust law from Calzetti et al. (2000) for stellar continuum and nebular emission. We assume that the stellar continuum reddening is proportional to the nebular reddening, (E(B-V) = $c~ \times$ E$_g$(B-V)), with the c value depending on the dust geometry of the ISM i.e., dust enhancement surrounding star forming regions.
 The rough linearity in the plot occurs because SFR(UV) is obtained from the observed $R$ band flux density. 
 We first assume Calzetti's constant $c = 0.44$, obtained from the study of local starburst galaxies. The median(E(B-V)) is 0.06 for the full sample. 
Some recent results suggest that dust properties at higher redshifts are different and a $c\sim1$ is required to explain the observations
 (\citealt{Erb2006b}). Therefore, we also plot $R$ versus reddening assuming $c=1.0$. In this case the median(E(B-V)) is 0.3 for the full sample. 
Comparing with the E(B-V) inferred by our SED fit (E(B-V)$_{\rm FULLsample}$ = 0.22[0.00-0.31], E(B-V)$_{\rm UV-faint}$ = 0.04[0.00-0.24], E(B-V)$_{\rm UV-bright}$ = 0.32[0.09-0.38]), presented in the plot as black triangles, we investigate which is the proper factor of proportionality between E(B-V) and Eg(B-V) in the framework of the Calzetti law. Under the assumption of simple radiative transfer of Ly$\alpha$ photons, Fig. 11 is compatible with any factor of proportionality $0.44<c<1.0$ at $z\simeq2.1$, but $c=1.0$ is disfavored at 68\% confidence level.
Uncertainties in Ly$\alpha$ radiative transfer make this comparison even less stringent, but a future independent determination of $c$ would allow this comparison to be used to constrain radiative transfer effects.

\subsection{X-ray stacking constraints on SFR and obscured AGN}
\label{sec:Xray}

In the central part of our field, the CDFS, we use the 4 Msec Chandra data to estimate the average stacked X-ray flux among 38 FULLsample objects.
We obtain an upper limit in the Chandra soft-band flux and a $>2\sigma$ detection in the hard band. The rest-frame hard X-ray flux is sensitive to the presence of high-mass X-ray binaries (HMXBs) in the galaxies. In this case 
In this case the involved time scale is 10-100 Myr, in which O and B stars compose the binary systems. For longer periods of time, low-mass X-ray binaries (LMXBs) and thermal emission can also contribute to the total hard X-ray budget, even if at $z\sim2$ the number of LMXBs is negligible (\citealt{Ghosh2001}). The HMXBs trace star formation, so we use these X-ray data as another way to infer the SFR (see also \citealt{Kurczynski2010}).  The 4 Msec data are so deep that only very-low luminosity or heavily-obscured AGN would escape individual detection.  As the observed-frame soft band could be obscured, we concentrate on the observed-frame hard band to infer the SFR. 

We use the \citealt{Persic2004} indicator to turn the hard X-ray luminosity into SFR$_{Persic}$. They assume that $f$ = (HMXB luminosity)/(total hard X-ray Luminosity) equal to 1 for high redshift systems implies that the hard X-ray luminosity is entirely from the HMXBs.
We also use the \citealt{Lehmer2010} equation that takes into account the contribution of LMXBs together with that of HMXBs to estimate the SFR, L$_{X-ray}$ = L(LMXB) $+$ L(HMXB) = $\alpha ~$M$_{*}+\beta ~$SFR). This way they fit the data with less scatter than Persic across a wider range of SFRs and X-ray luminosities.
When neglecting the factor involving the low-mass X-ray binary luminosity, the Lehmer and Persic equations are consistent. Our LAEs are characterized by low values of M$_{*}$, making the ``$\alpha ~$M$_{*}$'' term negligible with respect to the ``$\beta ~$SFR'' term.
In Tab. \ref{tab:Xray} we present stacked observed-frame soft and hard X-ray fluxes, and the inferred SFRs. The errors are estimated by measuring the background in $\sim$ 30 apertures around the central region, with the same size as for the source. Then, we calculate the mean and standard deviations of those background measurements. Obtaining an upper limit means that the stacked signal is not detected at the 2$\sigma$ level.  
The observed-frame soft band X-ray flux is $\leq 2.2$E-18 erg sec$^{-1}$ cm$^{-2}$, while the observed-frame hard band X-ray flux is ($4.6\pm1.9$)E-17 erg sec$^{-1}$ cm$^{-2}$.  The inferred  SFR$_{Persic}$ is $1180\pm480$ M$_{\odot}$ yr$^{-1}$, and SFR$_{Lehmer}$ is $730\pm300$ M$_{\odot}$ yr$^{-1}$, by assuming $\alpha=0$ and $\beta = (1.62 \pm 0.22)$ E+39 erg sec$^{-1}$ (M$_{\odot}$/yr)$^{-1}$.

Our SED fitting analysis shows that the typical LAE has an instantaneous SFR of 35[0.003-170] M$_{\odot}$ yr$^{-1}$ and the averaged $<$SFR$>_{100}$ is 4[2-30] M$_{\odot}$ yr$^{-1}$. Over the last 100 Myr and for a CSF assumption, SFR is directly comparable to the rest-frame UV flux density. Taking into account the dust amount through the Calzetti et al. (2000) law, we derive SFR$_{corr}$(UV). Based on the E(B-V) value measured in the SED fitting using a CSF, SFR$_{corr}$(UV) turns out to be 15[6-30] M$_{\odot}$ yr$^{-1}$ and $<$SFR$>_{100}$ is 4[2-9] M$_{\odot}$ yr$^{-1}$. Hence, for the FULLsample, SFRs of $\sim$10 M$_{\odot}$ yr$^{-1}$ seem to be derived by $<$SFR$>_{100}$, SFR$_{corr}$(UV), and instantaneous SFR estimations, independently of the SFH, but the SFR$_{hard~X-ray}$ is significantly higher.

There are a few plausible explanations for this discrepancy. We could be seeing a diffuse thermal plasma in X-rays. \citeauthor{Persic2002} (2002) discuss the diffuse component of X-ray luminosity in the context of X-ray SFR calibration, specifically.  The discrepancy could also be due to a contamination by a small number of low-luminosity or heavily obscured AGNs in our sample. They would dominate the averaged X-ray stack, but the median approach we are using in the SED analysis is not sensitive to them.
We look at the hardness ratio (HR=(counts$_{Xray-hard}$-counts$_{Xray-soft}$)/(counts$_{Xray-hard}$+counts$_{Xray-soft}$)) and at the ratio between X-ray and $R$ band fluxes to try to identify a discriminator between AGNs and pure star-forming galaxy imprints in our X-ray stack. Looking at Fig. 5 of Treister et al. (2009) and Fig. 8 of Wang at al. (2004), HR $\geq$ 0.43 and log(F$_{Xray-hard}$/F$_R) = -0.8\pm0.4$ values are consistent with having a mixed population of obscured AGNs and pure star-forming galaxies in our LAE sample.

\section{CONCLUSIONS}
\label{sec:conclusion}

\begin{figure}[!ht]
\begin{center}
\includegraphics[width=80mm]{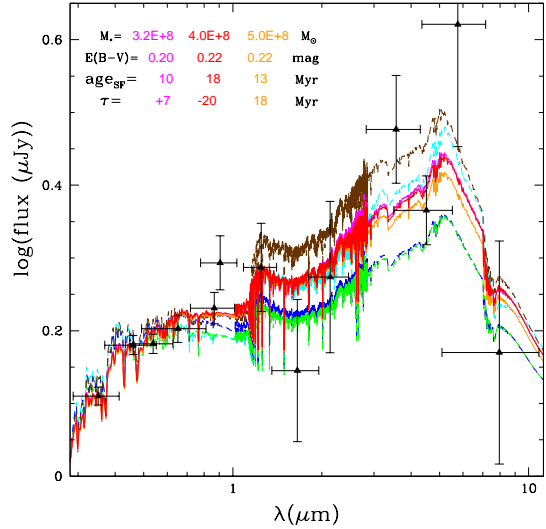}
\caption{Stacked SED of the FULLsample of $z\simeq2.1$ LAEs (black triangles with error bars). In solid lines we show the spectral energy distribution corresponding to the exponentially decreasing (magenta), exponentially increasing (red), and constant (orange) star formation 
histories, generated using the best fit parameters reported inside the plot. Dashed lines illustrate four models at the edge of the 68\% confidence level. The fit parameters for these models are M$_*$=8.91 $\times$ 10$^8$ M$_{\odot}$, E(B-V)=0.07, age$_{\rm SF}$=0.2273 Gyr, $\tau$=-1.39 Gyr (cyan); M$_*$=10$^9$ M$_{\odot}$, E(B-V)=0.1, age$_{\rm SF}$=0.114 Gyr, $\tau$=0.118 Gyr (brown); M$_*$=2.82 $\times$ 10$^8$ M$_{\odot}$, E(B-V)=0.16, age$_{\rm SF}$=0.018 Gyr, $\tau$=-4.0 Gyr (blue); M$_*$=2.82 $\times$ 10$^8$ M$_{\odot}$, E(B-V)=0.17, age$_{\rm SF}$=0.018 Gyr, $\tau$=2.812 Gyr (green).}
\label{fullsamplebestfits}
\end{center}
\end{figure}

\begin{figure}[!ht]
\begin{center}
\includegraphics[width=140mm]{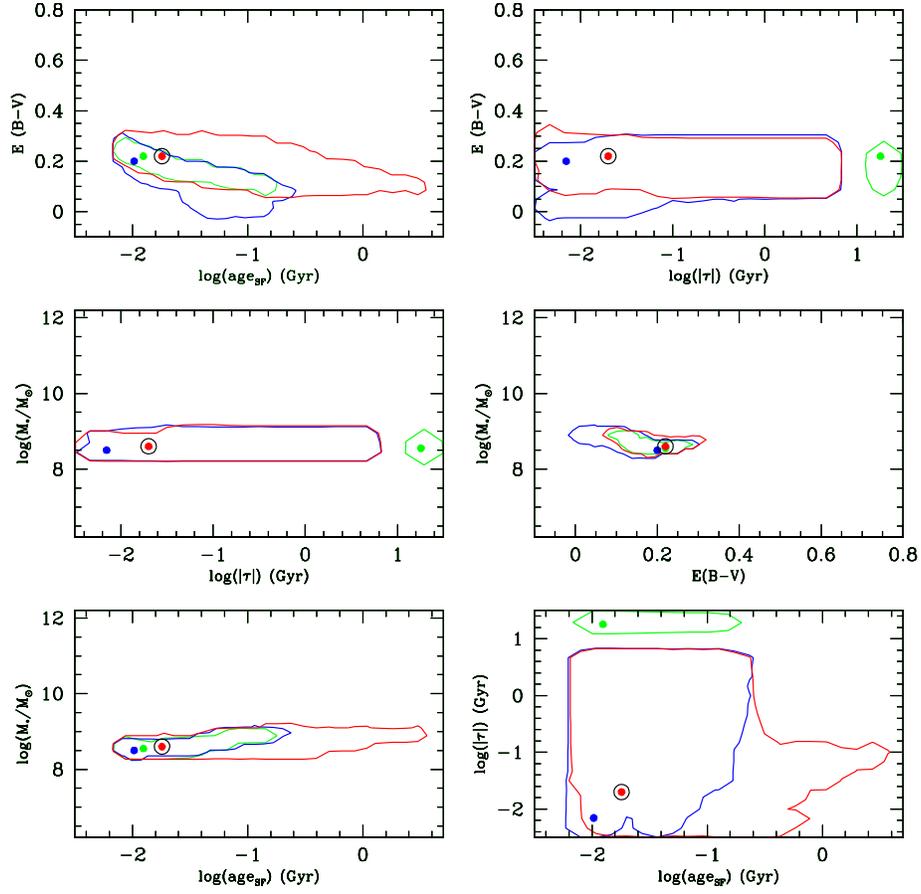}
\caption{68\% confidence levels for the model parameters in the case of the exponentially decreasing (blue), exponentially increasing (red), and constant (green) SFHs. Different panels show different combinations of parameters. The contours show that while M$_*$ and E(B-V) are well constrained, a wide range of ages and even wider range of $\tau$ are allowed at 68\% confidence.
Fits of comparable quality can be obtained for all three SFHs.
The best fit models are indicated as colored dots. The dot with the black circle represents the best fit in the entire parameter space.}
\label{6panel}
\end{center}
\end{figure}

\begin{figure}[!ht]
\begin{center}
\includegraphics[width=150mm]{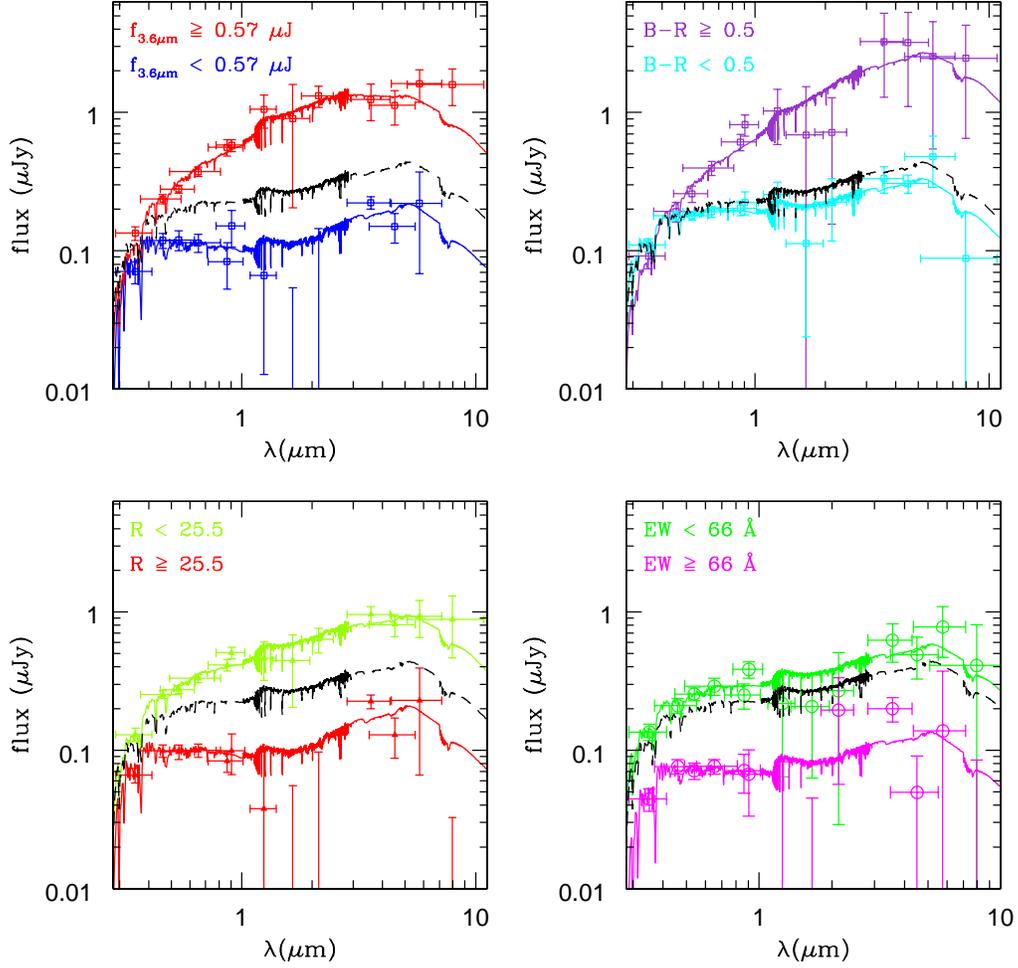}
\caption{Observed median SEDs (color symbols) and best fit model SEDs (color curves) in a log-log plot. The FULLsample best fit SED is plotted (black dashed curve) for comparison. Each panel shows pairs of sub-samples as labelled.}
\label{SEDpanel}
\end{center}
\end{figure}

\begin{figure}[!ht]
\begin{center}
\includegraphics[width=140mm]{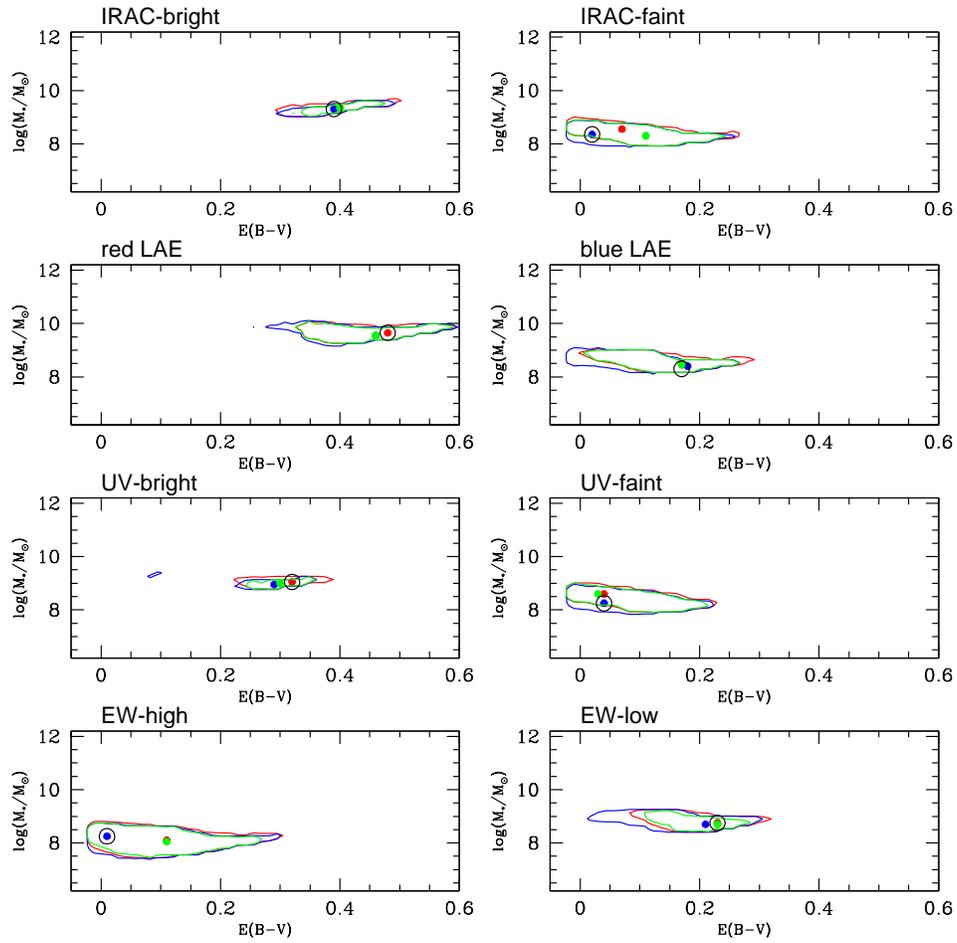}
\caption{68\% confidence level in Log(M$_*$/M$_{\odot}$) versus E(B-V) plane for the exponentially decreasing (blue), exponentially increasing (red), and constant (green)
SFHs for different LAE sub-samples. The dots represent the best fits for each of the three SFHs. 
The dot with the black circle represents the best fit in the entire parameter space. These parameters are well constrained for every combination of sub-samples and SFHs. More massive sub-samples are characterized by higher dust reddening.}
\label{contsubsample}
\end{center}
\end{figure}

\begin{figure}[!ht]
\begin{center}
\includegraphics[width=140mm]{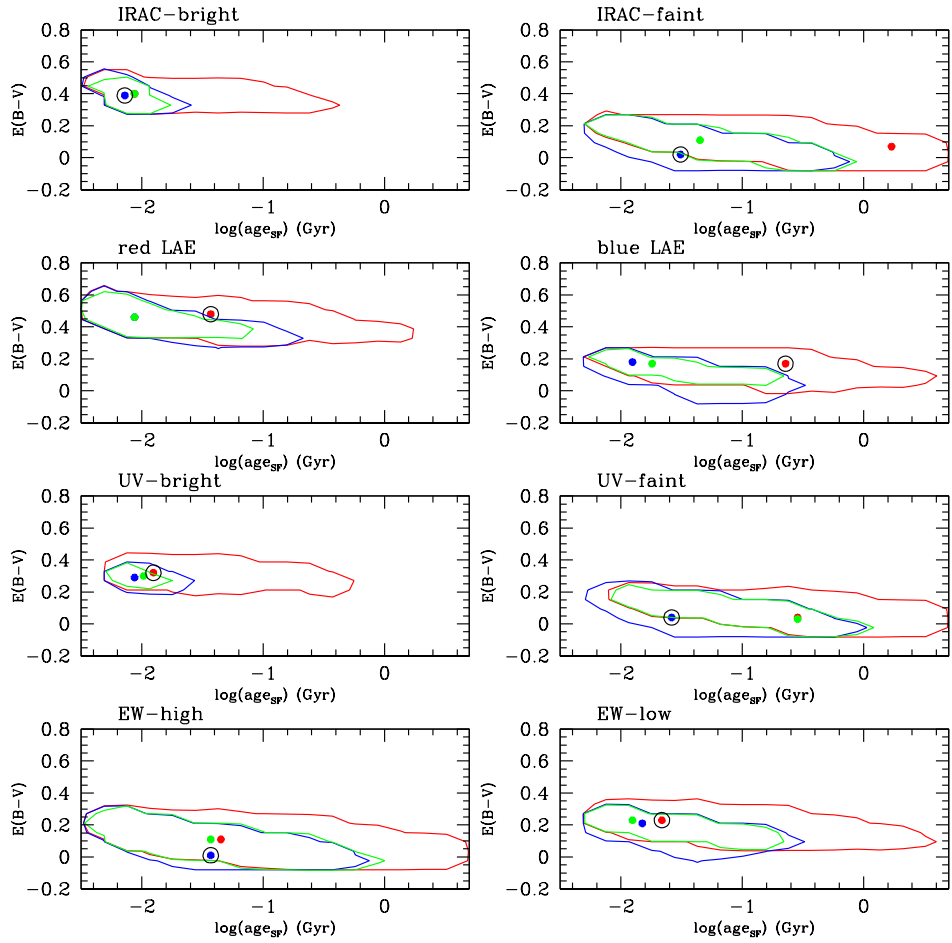}
\caption{68\% confidence level in E(B-V) versus log(age$_{\rm SF}$) plane for the three SFHs, with the symbols and colors as in Fig. \ref{contsubsample}. 
We see minimal degeneracy between these two parameters.  age$_{\rm SF}$ is not well constrained except that IRAC-bright and UV-bright sub-samples have ages less than 20 Myr in the exponentially declining and constant SFHs.}
\label{contageE}
\end{center}
\end{figure}

\begin{figure}[!ht]
\begin{center}
\includegraphics[width=140mm]{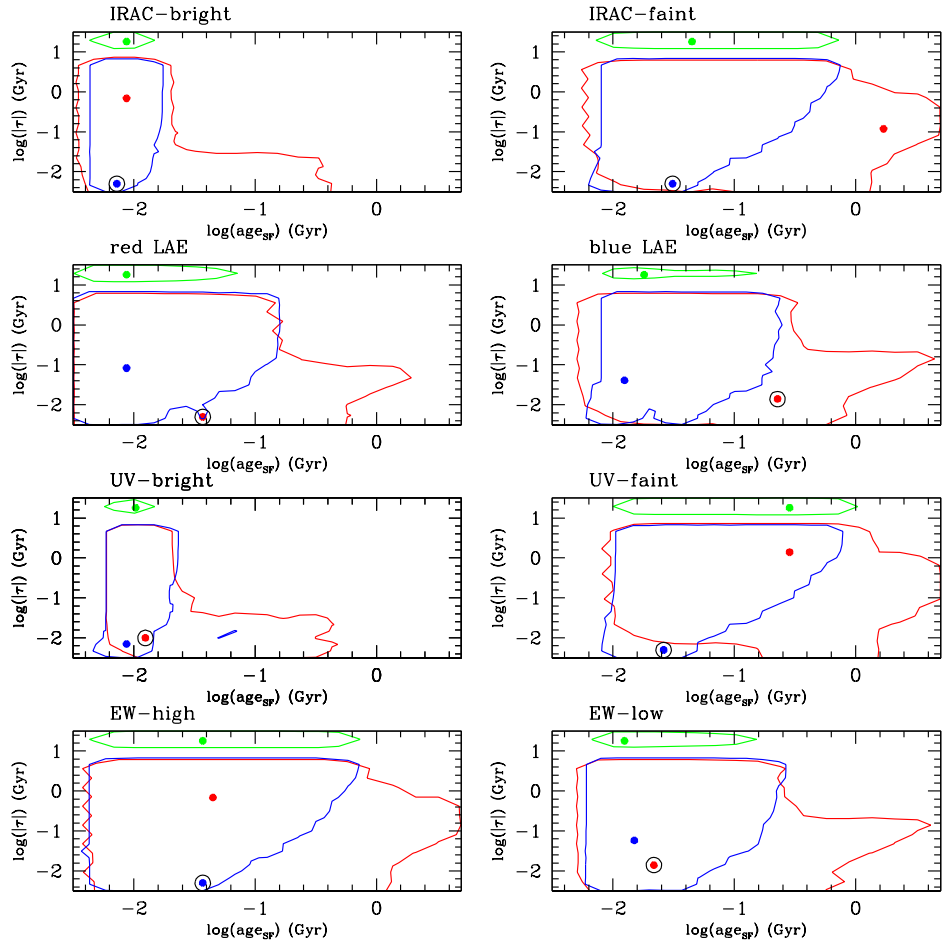}
\caption{68\% confidence level in Log($|\tau|$) versus log(age$_{\rm SF}$) plane for the three SFHs, with symbols and colors as in Fig. \ref{contsubsample}. The contours show that even if age$_{\rm SF}$ and $\tau$ are not well constrained individually, some regions of the parameter space are disfavored at the 68\% confidence level.}
\label{conttauage}
\end{center}
\end{figure}

In this work we calculated the observed spectral energy distribution of the $z\simeq2.1$ LAE sample and its sub-samples (Tab. \ref{tab:subsampledefinition}), obtained based upon photometric properties by dividing the sample roughly in halves. 
We mainly concentrated on the IRAC-bright and faint sub-samples, because the separation at observed 3.6 $\micron$ is expected to be a separation based on the stellar mass; on the red-LAEs as a special category with observed SED different from the bulk of $z\simeq2.1$ LAEs; the UV-bright and UV-faint sub-samples, because the separating $R$ magnitude represents the rest-frame UV continuum of our galaxies.
The complete $z\simeq2.1$ LAE sample, defined in  Gu10 and here in \S 2, is used to investigate the EW correlation with SED parameters. 
To allow the properties of our LAE sample to represent the general properties of LAEs at this redshift in the Universe, we take sample variance into account in our error estimation by using bootstrapping technique (\S 2).
We used the CB10 stellar population code to generate continuum SED models, probing a wide range of star formation histories parametrized by the e-folding time, $\tau$. We assumed the Salpeter IMF to be able to compare with results in the literature. It was also justified by the fact that the stacked flux density in the NB5015 produced a non-detection (\S \ref{sec:model}). In fact at $z\simeq2.1$ that narrow-band flux would reveal the HeII emission, imprint of top heavy IMF and/or very-low metallicity stellar population. The EW(HeII $\lambda 1640$), implied by NB5015 non detection, is consistent with zero, as it may be expected for a normal IMF galaxy.  
We looked for the best fit SED (Fig. \ref{fullsamplebestfits}, Fig. \ref{SEDpanel}) and the 68\% confidence level of fitting parameters of the $typical$ LAEs of our sample and sub-samples. In Fig. \ref{6panel}, \ref{contsubsample}, \ref{contageE}, and \ref{conttauage} we presented joint contours of different combinations of parameters and star formation histories. Equally good fits of our data can be obtained assuming exponential increasing, decreasing or constant star formation rates.
The stellar mass, M$_*$, and the reddening, E(B-V), are well constrained, while age$_{\rm SF}$ has large uncertainties and $\tau$ is not constrained at all.

${\it{Heterogeneity}}$ can be revealed in a sample of galaxies either by fitting individual galaxies, if there is enough signal-to-noise
(e.g., \citealt{Fin:2008}), or by our approach of stacking sub-samples selected according to their observed photometric properties. 
Our results are in agreement with Nilsson et al. (2009, 2011). 
The Gu10 clustering result showed that $z\simeq2.1$ LAEs evolve into local Universe galaxies with median luminosity equal to L$^*$ and that $z\simeq2.1$ LAEs therefore represent  the building blocks of $z\sim0$ Milky Way-type galaxies. 
From our median statistics analysis, our results show that these LAEs appear characterized by median M$_*\lesssim 10^9$ M$_{\odot}$ and a $\sim$ 40 M$_{\odot}$ yr$^{-1}$ median instantaneous SFR. This moderate-to-high SFR value might be caused by rapid accretion of mass, which, together with merging phenomena, must play a major role in causing LAEs to evolve into $z\sim0$ L$^*$ galaxies. The observed starburst is presumably not the only phase of active star formation happening inside these galaxies
between formation and the present day; multiple episodes of SF are likely to contribute to the growth of stellar mass. We could also be observing Ly$\alpha$ emission as the effect of the SF in particular locations within a galaxy where the Ly$\alpha$ photons are free to propagate, even if other galaxy regions are temporarily inactive or dustier.   

The main results obtained from SED fitting and further discussions are the following:

$\bullet$  $z\simeq2.1$ LAEs appear to be low mass galaxies with moderate amounts of dust. The instantaneous SFR values are estimated of the order of tens of M$_{\odot}$ yr$^{-1}$, with an extreme value of 170 M$_{\odot}$ yr$^{-1}$ within the 68\% confidence level. However, the star formation rate averaged over the last 100 Myr,  $<$SFR$>_{100}$, is consistent with $\sim$ 4 M$_{\odot}$ yr$^{-1}$. The mean stellar age, t$^*$, is $\sim$ 10 Myr, so that the LAE SED is consistent with being dominated by young stars i.e., LAEs are undergoing a period of intense star formation. 

$\bullet$ $z\simeq2.1$ LAEs appear in median dustier (E(B-V) = 0.22[0.09-0.28]) than $z\simeq3.1$ LAEs assuming constant star formation rate model, but present consistent stellar masses. 

\begin{figure}[!ht]
\begin{center}
\includegraphics[width=80mm]{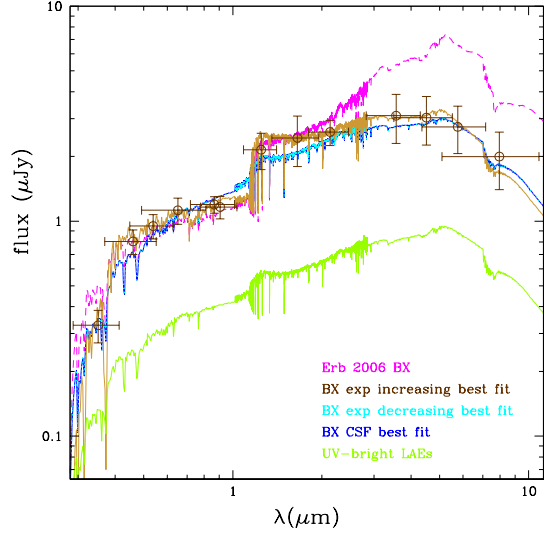}
\caption{Observed SED (brown circles with error bars) for the sample of spectroscopically confirmed BX star forming galaxies at $2.0<z_{spec}<2.2$.  The solid brown curve shows our best fit SED model, obtained with an exponentially decreasing SFH, while the solid blue and dotted cyan curves show the best fit obtained assuming constant and exponentially decreasing SFR. The dashed magenta line shows the SED which corresponds to the median best fit parameters of the sample from \citealt{Erb:2006}. As a comparison we show the best fit model SED of our UV-bright sub-sample in olive green.} 
\label{BX}
\end{center}
\end{figure}

$\bullet$ The typical $z\simeq2.1$ LAEs appear to be less massive (M$_*\leq$ 10$^8$ M$_{\odot}$) than  $z\simeq2.1$ UV-continuum selected star forming galaxies (Fig. \ref{BX}), implying that LAEs tend to occupy the low-mass end of the distribution of star forming galaxies at $z\sim2$.

$\bullet$ The IRAC-bright LAEs are in median significantly more massive and dustier than those that are faint in IRAC.
The $R<25.5$ LAEs are also more massive than the fainter ones. 
Our typical $z\simeq2.1$ LAE is blue with median($B-R)=0.13$, but the 15\% of the objects classified as red-LAEs are significantly more massive and dustier, with E(B-V)~$>0.3$. 

$\bullet$ The sub-sample of LAEs characterized by high-EW(Ly$\alpha$) shows the lowest stellar mass and a moderate dust amount (E(B-V) = [0.0-0.3]). 

\begin{figure}[!ht]
\begin{center}
\includegraphics[width=70mm]{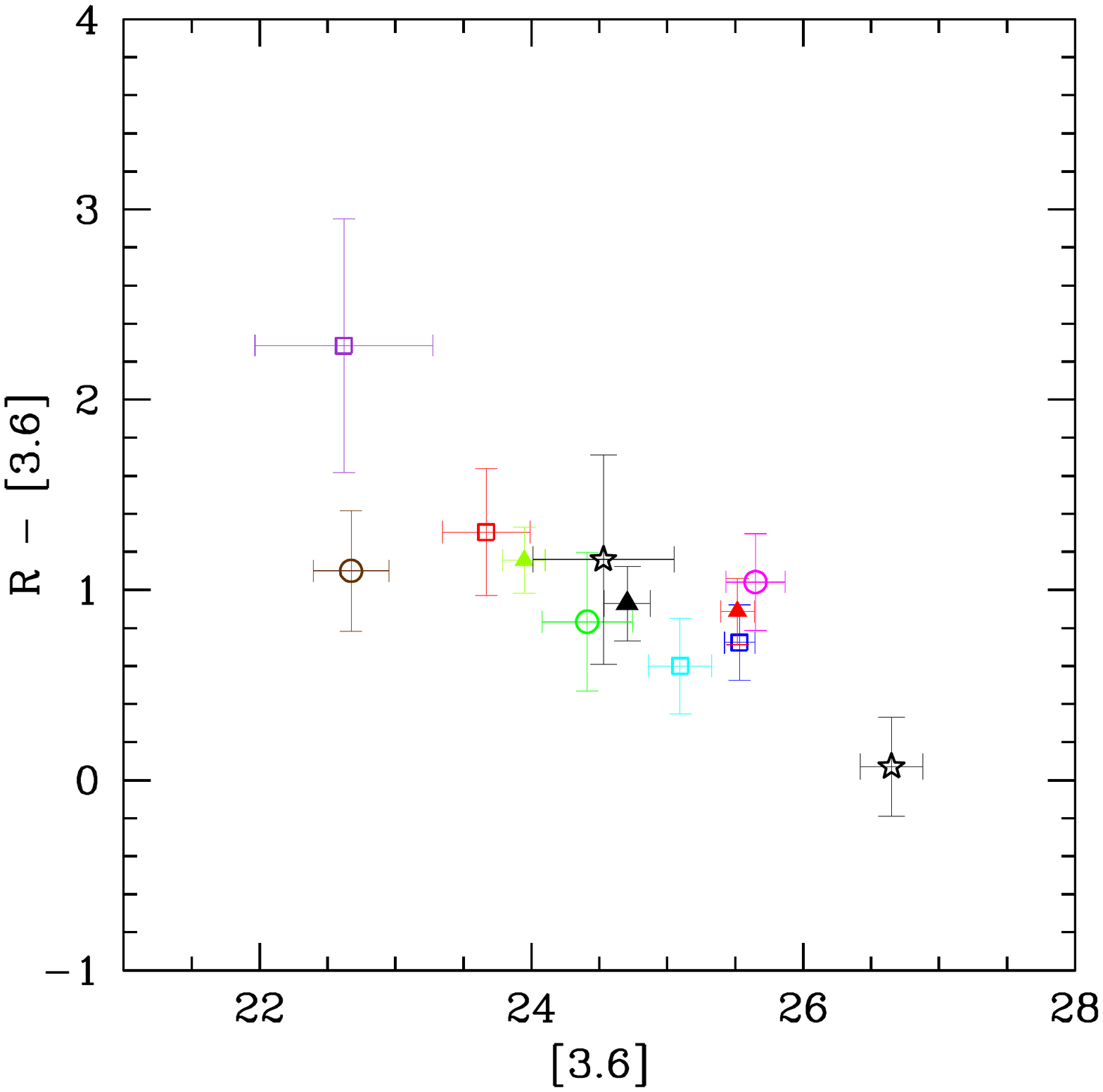}
\includegraphics[width=70mm]{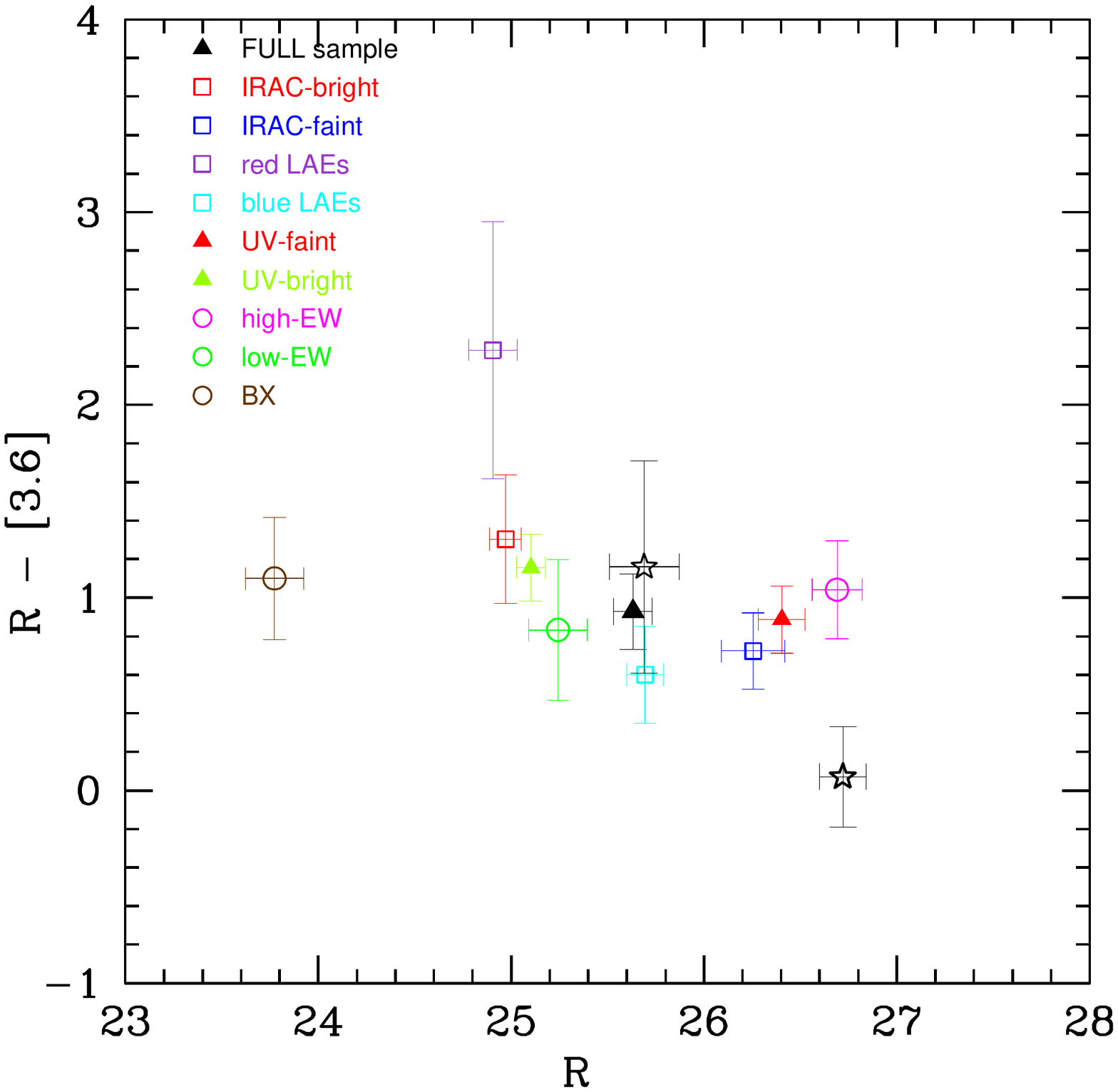}
\caption{Color-magnitude diagram for color-mass (left) and color-SFR (right) relations. The symbols represent the $z\simeq2.1$ LAE sub-samples with the color coding as in Fig. \ref{SEDpanel} (see legend). The black stars represent the results of Lai et al. (2008) for the $z\simeq3.1$ IRAC L08und (on the extreme right) and L08det LAE sub-samples. 
The larger horizontal error bars in the left panel are caused by larger flux uncertainties in the observed IR in comparison to the optical bands.} 
\label{R36R}
\end{center}
\end{figure}

$\bullet$ $z\simeq2.1$ LAEs have a wide range of stellar masses (Fig. \ref{R36R}). There is a continuum of properties from the redder and more massive to the bluer and less massive typical LAEs, which are the majority. This is another indication that LAEs are located at the lower mass end of the $z\sim2$ star forming galaxy mass function, where the bulk of galaxies lie.  

\begin{figure}[!ht]
\begin{center}
\includegraphics[width=70mm]{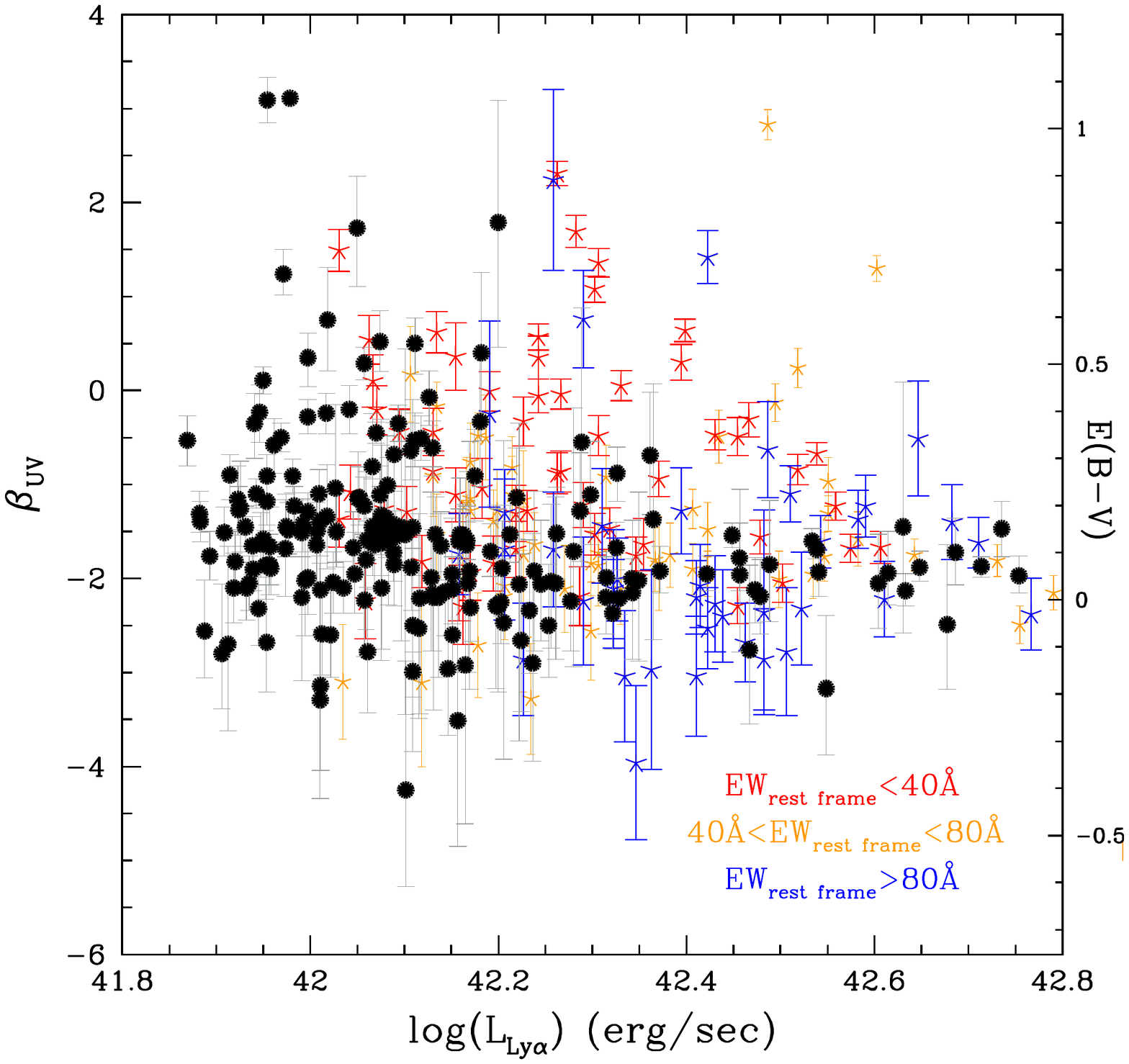}
\includegraphics[width=70mm]{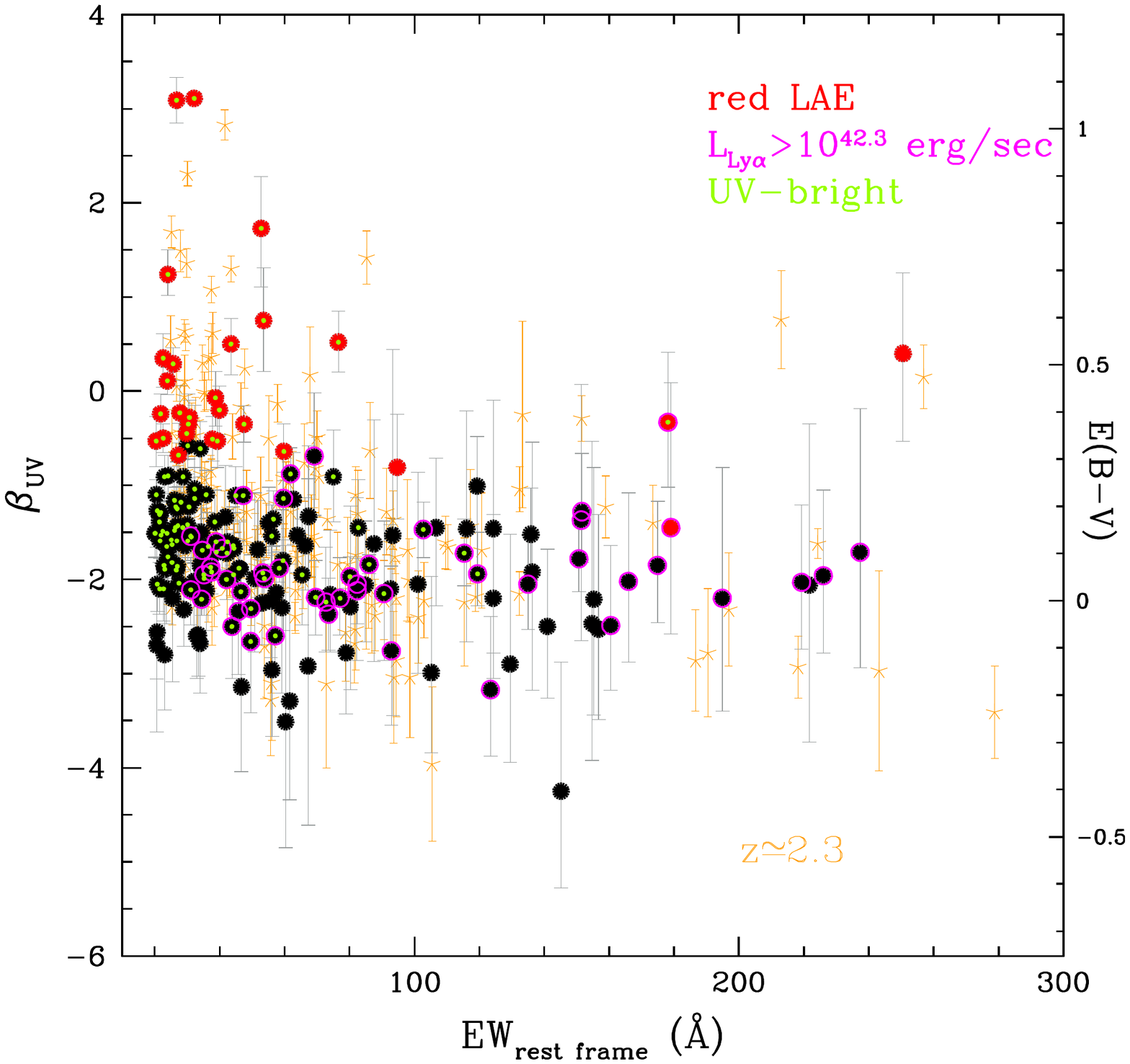}
\caption{Rest-frame UV slope ($\beta_{UV}$) versus L$_{Ly\alpha}$ (left) and rest-frame equivalent width (right) for $z\simeq2.1$ LAE samples. The y-axes also show the stellar continuum E(B-V) parameter from the \citealt{Meurer1999} formalism. Black dots with gray error bars correspond to the $z\simeq2.1$ LAEs. 
In the left panel we show the $z\simeq2.3$ data separating them based upon equivalent width: EW$_{rest-frame}<40$ {\AA} in red, 40 {\AA}~$<$~EW$_{rest-frame}<80$ {\AA} in orange, and EW$_{rest-frame}>80$ {\AA} in blue crosses.
In the right panel the orange crosses with error bars correspond to the $z\simeq2.3$ LAEs from Nilsson et al. (2009). Three $z\simeq2.1$ sub-samples are also presented here: red LAEs as red dots, LAEs with L$_{Ly\alpha}>10^{42.3}$ erg sec$^{-1}$ as magenta circles, and the UV-bright LAEs as small green dots.} 
\label{beta}
\end{center}
\end{figure}

\begin{figure}[!ht]
\begin{center}
\includegraphics[width=70mm]{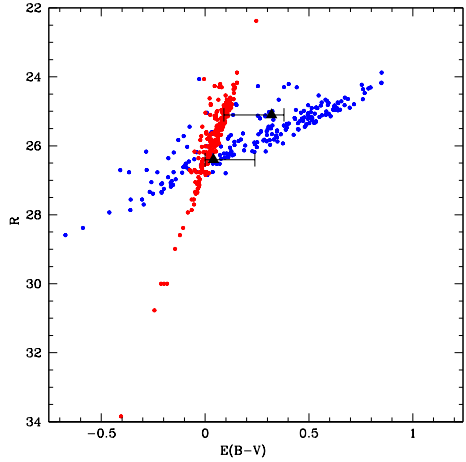}
\caption{Observed $R$ band magnitude versus stellar continuum reddening, E(B-V), inferred from the Ly$\alpha$/UV luminosity ratio and the assumption of simple Ly$\alpha$ radiative transfer. We assume the dust law from Calzetti et al. (2000) and E(B-V) = c $\times$ E$_g$(B-V). Red dots correspond to $c=0.44$ and blue dots to $c=1.0$. Black triangles represent the median observed $R$ magnitude and the best fit E(B-V) values from the SED analysis for the UV-bright and UV-faint sub-samples.} 
\label{RE}
\end{center}
\end{figure}

$\bullet$ The red-LAEs are characterized by EW$_{rest-frame}<50$ {\AA} (Fig \ref{beta}). The Ly$\alpha$ luminous LAEs are consistent with flat rest-frame UV spectra and can present equivalent widths up to 200 {\AA}.  

$\bullet$ The best-fit instantaneous SFR and $<$SFR$>_{100}$ differ significantly in several sub-samples, making it important to consider the timescales of the various SFR estimators when studying starburst galaxies, such as LAEs. 

Our results point out differences in the dust amount and SFR of LAEs between redshifts $z\simeq3.1$ and $z\simeq2.1$.
A possible explanation is that the redder and dustier LAEs, observed at the later time in the evolution of the Universe, are experiencing evolved phases of star formation, unlike their dust-free $z\simeq3.1$ counterparts (Ga07, L08, \citealt{Acq2011}). Future work will investigate whether these results are robust by using uniform SED fit methodology and stellar population synthesis modeling for the two epochs.
If the indications coming from this work and N10 are confirmed, this will be a key step in our understanding of how the building blocks of L$^*$ galaxies evolve with cosmic time.

\acknowledgments

We acknowledge Gustavo Bruzual and Stephen Charlot for providing the newest GALEXEV code; Kamson Lai, Guillermo Blanc, Mike Berry, John Feldmeier, Kim Nilsson, Jean Walker, and Zheng Zhen-Ya for useful discussion and comments on the paper.   
This project made use of 50,000 cpu hours in the Geryon computing cluster at the Centro de Astro-ingenier\'\i a UC.
We are grateful for support from  Fondecyt (\#1071006),
Fondap 15010003, Proyecto Conicyt/Programa de Financiamiento Basal para Centro Cient'ficos y Tecnol—gicos de Excelencia (PFB06),
Proyecto Mecesup 2 PUC0609, and the ALMA-SOCHIAS fund for travel grants.  
This material is based on work supported by NASA through an award issued by JPL/Caltech, by 
the National Science Foundation under grants AST-0807570 and AST-0807885, and by the Department of 
Energy under grants DE-FG02-08ER41560 and DE-FG02-08ER41561.   
Support for the work of E.T. was provided by the National Aeronautics and Space Administration through Chandra Post-doctoral Fellowship Award Number PF-90055 issued by the Chandra X-ray Observatory Center, which is operated by the Smithsonian Astrophysical Observatory for and on behalf of the National Aeronautics Space Administration under contract NAS8-03060. 
Support for the work of KS was provided by NASA through Einstein Postdoctoral
Fellowship grant number PF9-00069 issued by the Chandra X-ray Observatory
Center, which is operated by the Smithsonian Astrophysical Observatory for and
on behalf of NASA under contract NAS8-03060. 
E.G. thanks the Department of Physics at U.C. Davis for hospitality during the completion of this research.  
L.G thanks Rutgers University for hosting her during collaborative research. 

{\it Facilities:} \facility{Blanco}

\clearpage

\begin{deluxetable*}{ccccccccccccc}
\tabletypesize{\scriptsize}
\tablecaption{Limiting magnitude in the 13 MUSYC broad bands in ECDF-S}
\tablewidth{0pt}
\tablehead{
\colhead{\bf{U}}      &   \colhead{\bf{B}}      &
\colhead{\bf{V}}        &   
\colhead {\bf{R}} &   \colhead{\bf{I}}      &    \colhead{\bf{z}}      &   \colhead{\bf{J}}      &  \colhead{\bf{H}}      &  \colhead{\bf{K}}      &  \colhead{\bf{3.6$\mu$m}}   &  \colhead{\bf{4.5$\mu$m}}      &  \colhead{\bf{5.8$\mu$m}}      &    \colhead{\bf{8.0$\mu$m}} }
\startdata
26.1 & 26.9 & 26.5  & 26.5 & 24.6 & 23.6 & 22.8 & 22.3 & 21.5 & 23.9 & 23.7 & 21.9 & 21.8 
\enddata
\label{tab:limitingMAG}
\end{deluxetable*}

\begin{deluxetable*}{|l|c|c|}
\tabletypesize{\scriptsize}
\tablecaption{Sub-sample definitions}
\tablewidth{0pt}
\tablehead{
\colhead{\bf{Sub-sample name}}      &   \colhead{\bf{Selection}}      &
\colhead{\bf{N$_{LAE}$}}  }
\startdata
FULLsample & entire $z\simeq2.1$ LAE sample &  216 \\
\hline
IRAC-bright & f$_{3.6\mu m}\geq0.57 \mu J$ & 47 \\
IRAC-faint & f$_{3.6\mu m}<0.57 \mu J$ &  67 \\
\hline
red-LAE & $B-R\geq0.5$  & 34 \\
blue-LAE  & $B-R<0.5$ &  182\\
\hline
UV-bright  & $R<25.5$ & 118 \\
UV-faint  & $R\geq25.5$ & 98  \\
\hline
 & log(L$_{Ly\alpha})\geq42.1$ &  \\
high-EW  & EW$_{rest-frame}\geq$66 {\AA} & 60 \\
   & log(L$_{Ly\alpha})\geq42.1$ &  \\
low-EW & EW$_{rest-frame}<$66 {\AA} & 59
\enddata
\tablecomments{The number of LAEs in each sub-sample is based on having coverage in the optical bands. The only exception is for the sub-samples defined based on IRAC 3.6 $\mu$m, which are selected from the 114 LAEs in the IRAC ``clean'' regions.}
\label{tab:subsampledefinition}
\end{deluxetable*}

\begin{deluxetable*}{|c|c|c|c|c|c|c|c|c|}
\tabletypesize{\tiny}
\tablecaption{Overlap of objects among the sub-samples}
\tablewidth{0pt}
\tablehead{
\colhead{sub-sample} & \colhead{IRAC-bright}    &  \colhead{IRAC-faint}  & \colhead{red-LAEs} & \colhead{blue-LAEs} &  \colhead{UV-bright} & \colhead{UV-faint}  & \colhead{EW-high} & \colhead{EW-low} }      
\startdata
IRAC-bright&{\bf{47}}&0&16& 31 &42&5& 4& 14 \\
IRAC-faint&0& {\bf{67}}& 5 & 62 &16& 51&30& 14\\ 
\hline
red-LAEs &16&5&{\bf{34}}& 0 & 25&9& 8&6 \\
blue-LAEs & 31& 62& 0& {\bf{182}}& 93& 89&52& 53 \\
\hline
UV-bright& 42& 16&25& 93 &{\bf{118}}&0& 5& 49 \\
UV-faint& 5& 51& 9 & 89 & 0& {\bf{98}}&55& 10 \\
\hline
EW-high&4& 30& 8& 52 &5& 55& {\bf{60}}&0\\
EW-low&14& 14& 6& 53 & 49& 10&  0& {\bf{59}}
\enddata
\tablecomments{For each sub-sample we report the number of objects belonging to other sub-samples. Boldface numbers indicate the actual number of objects in each sub-sample.}
\label{tab:overlap}
\end{deluxetable*}

\begin{deluxetable*}{cccc}
\tabletypesize{\scriptsize}
\tablecaption{Free parameters and value ranges used in the SED fitting process}
\tablewidth{0pt}
\tablehead{
\colhead{\bf{parameter}}      &   \colhead{\bf{Min. Value}}      &
\colhead{\bf{Max. Value}}        &   
\colhead {\bf{step}}  }
\startdata
Parameter & Min. Value & Max. Value  & step \\
\hline
E(B-V) (mags) & 0.00 & 0.60 & 0.01 \\
\hline
log(M$_*$/M$_\odot$) & 6.00 & 12.00 & 0.05 \\
\hline
 age$_{\rm SF}$(Gyr) & 0.005 & 3.00 & $\sim$0.1 dex \\
\hline
$\tau$(Gyr) & -4 & 18 & $\sim$0.15 dex
\enddata
\label{tab:freeparam}
\end{deluxetable*}

\begin{deluxetable*}{|l|c|c|c|c|c|c|c|c|c|}
\tabletypesize{\scriptsize}
\tablecaption{Full sample SED parameters}
\tablewidth{0pt}
\tablehead{
\colhead{\bf{SFH}}      &    \colhead{\bf{log(M$_*$/M$_{\odot}$)}}    &
    \colhead{\bf{E(B-V)}}     &   
\colhead {\bf{age$_{\rm SF}$(Gyr)}} &   \colhead{\bf{$\tau$(Gyr)}} &  \colhead{\bf{SFR (M$_\odot$/yr) }} & \colhead{\bf{$<$SFR$>_{100}$ (M$_\odot$/yr)}} &  \colhead{\bf{$\chi^2_{best~fit}$/d.o.f}}  }
\startdata
all & 8.6[8.4-9.1] & 0.22[0.00-0.31] & 0.018[0.009-3]  & -0.02[all] & 35[0.003-170] &  4[2-30] &12.3/9 \\
$\tau>0$ & 8.5[8.4-9.2] & 0.20[0.00-0.29] & 0.010[0.009-0.227]  & 0.007[all] & 14[0.003-60] & 3[2-15] &  12.4/9\\
$\tau<0$ & 8.6[8.5-9.0] & 0.22[0.07-0.31] & 0.018[0.009-3]  & -0.020[all] & 35[5-170] & 4[2-30] &  12.3/9\\
CSF & 8.6[8.4-9.0] & 0.22[0.09-0.28] & 0.012[0.009-0.161]  & $\infty$ & 30[6-60]  & 4[2-9] &  12.6/10
\enddata
\tablecomments{M$_*$ is the observed stellar mass, taking into account the gas reprocessing inside a galaxy. age$_{\rm SF}$ is the age since star formation began. SFR is derived from the star formation history as $\frac{N}{|\tau|}$exp(-$\frac{age_{SF}}{\tau})$ (Equation \ref{sfrh}). 
For the constant star formation model (CSF) we fix $\tau$ to be an arbitrarily high value of 18 Gyr in our code.}
\label{tab:solofull}
\end{deluxetable*}
 
\begin{deluxetable}{ccccccc}
\tabletypesize{\scriptsize}
\tablecaption{$z\simeq2$ and $z\simeq3$ SED fitting}
\tablewidth{0pt}
\tablehead{
\colhead{\bf{reference}}      &   \colhead{\bf{z}}      &
\colhead{\bf{$\tau$(Gyr)}}        &
\colhead{\bf{E(B-V)}}          &      \colhead{\bf{log(M$_*$/M$_\odot$)}}    &
\colhead{\bf{age$_{\rm SF}$(Gyr)}}   &    {\bf{SFR(M$_{\odot}$/yr)}}  }
\startdata
N07 & 3.15 & $\infty$ & 0.07[0.02-0.09] &  8.67[8.18-8.95] &  0.85[0.43-0.98] & 0.66[0.35-1.17]\\
Ga07 & 3.1 & 0.75$^{+0.25}_{-0.25}$& 0.00[0.00-0.08] &  9.00[8.80-9.20] &  0.02[0.01-0.05] & 2.00[1.00-3.00]\\
O10und &3.1 & $\infty$ & 0.03[0.00-0.08] &  8.12[7.75-8.38] &  0.07[0.01-0.18] & 0.36[0.19-0.78] \\
O10det &3.1 & $\infty$ & [0.00-0.70] &  [8.97-10.43] &  [0.005-0.407] & [11-5600] \\
N10 & 2.3 & 2 SSP & 0.07[0.03-0.20] & 9.83[9.66-9.98] & 0.07[0.02-0.09] & - \\
FULLsample (CSF) & 2.1 & $\infty$ &  0.22[0.09-0.28]  & 8.6[8.4-9.0] & 0.012[0.009-0.161] & 30[6-60]  \\
FULLsample ($\tau>0$) & 2.1 & 0.007[all]  &0.20[0.00-0.29]  & 8.5[8.4-9.2] & 0.010[0.009-0.227] & 14[0.003-60] \\
FULLsample ($\tau<0$) & 2.1 & -0.020[all]  & 0.22[0.07-0.31] & 8.6[8.5-9.0] &0.018[0.009-3]  & 35[5-170] \\
\hline
IRAC-bright & 2.1 & $\infty$ & 0.40[0.32-0.46] & 9.4[9.2-9.5] & 0.009[0.006-0.015] & 260[110-530] \\
L08det & 3.1 & $\infty$ & 0.00[0.00-0.10]  & 9.95[9.78-10.08]  & 1.60[1.20-2.00] & 6[5-7] \\
\hline
IRAC-faint & 2.1 & $\infty$ & 0.11[0.00-0.25] & 8.3[8.0-8.9] & 0.045[0.009-0.716] & 5[1-25]\\
L08und &  3.1 & $\infty$ & 0.00[0.00-0.10] & 8.48[8.00-8.85] & 0.16[0.05-0.30] & 2[1-3]\\
\hline  
BX & $2.0<z_{spec}<2.2$ & $\infty$ & 0.35[0.31-0.37] & 9.6[9.5-9.7] & 0.009[0.009-0.01]  & 470[315-526]  
\enddata
\tablecomments{The references indicate Nilsson et al. (2007)=N07,
Nilsson et al. (2011)=N10, \citeauthor{Ono2010} (2010)=O10 with $K(3\sigma$)=24.1, Gawiser et al. (2007)=Ga07 with mag$_{AB}$ [3.6]~$>25.2$ and a 2-population model. CSF indicates constant star formation. L08det, L08und are IRAC-detected and IRAC-undetected subsamples from \citeauthor{Lai:2008} (2008). FULLsample, IRAC-bright, IRAC-faint, and BX are from this work.}
\label{tab:ALLcomp}
\end{deluxetable}

\clearpage
\begin{turnpage}
\begin{deluxetable}{|cccccccccr|}
\tabletypesize{\tiny}
\setlength{\tabcolsep}{0.02in}
\tablecaption{Full parameter space allowed parameters for the $z\simeq2.1$ LAE subsamples}
\tablewidth{0pt}
\tablehead{
\colhead{\bf{subsample}}      &   \colhead{\bf{log(M$_*$)}}       &
   \colhead{\bf{E(B-V)}}    &
 \colhead{\bf{$\tau$}}    & \colhead{\bf{age$_{\rm SF}$}}          &     
\colhead{\bf{t$^*$}}        &
\colhead{\bf{SFR}} &  \colhead{\bf{$<$SFR$>_{100}$}} &  \colhead{\bf{age$_{\rm SF}$/$\tau$}} &\colhead{\bf{$\chi^2_{best}$ [$\chi^2_{68\%}$]}}}
\startdata
 FULLsample &  8.6[8.4-9.1] & 0.22[0.00-0.31] & -0.02[all] & 0.018[0.009-3] &  0.007[0.004-0.85] & 35[0.003-170]  & 4[2-30] &   -0.9[(-100)-11]  &12.3[$\leq$21.9]\\
 \hline
 IRAC-bright &  9.3[9.1-9.7] & 0.39[0.30-0.50] & 0.005[all] & 0.007[0.006-0.404] &    0.005[0.002-0.16] & 130[40-1310]  & 21[10-120] &   1.5[(-64)-2]  &12[$\leq$21.4]\\
 IRAC-faint &  8.4[7.9-8.9] & 0.02[0.00-0.30] & 0.005[all] & 0.031[0.007-3] &      0.03[0.003-1.5] & 0.1[0.0003-83]  & 3[1-15] &  6.2[(-227)-13]  & 21.8[$\leq$38.9]\\
 \hline
 red-LAEs &  9.7[9.2-10.0] & 0.48[0.25-0.56] & -0.005[all] &  0.04[0.005-2.1] &   0.003[0.002-0.57] & 1090[1-2880]  & 55[20-340] &  -7.4[(-80)-7.4]  & 7.8[$\leq$ 13.9]\\
 blue-LAEs &  8.5[8.3-9.0] & 0.17[0.00-0.25] & -0.014[all] & 0.227[0.009-3] &     0.01[0.004-0.85] & 25[0.02-110]  & 3[2-20] & -16[(-100)-9]  & 6.2[$\leq$11.0]\\
 \hline
  UV-bright &  9.1[8.9-9.4] & 0.32[0.09-0.38] & -0.01[all] & 0.012[0.007-0.404] &  0.005[0.003-0.16] & 160[1-420]  & 10[6-70] &   -1.2[(-64)-5]  & 9.1[$\leq$ 16.2]\\
 UV-faint &  8.3[7.9-9.0] & 0.04[0.00-0.24] & 0.005[all]  & 0.026[0.009-3] &   0.02[0.005-1.5] & 0.2[0.002-64]  & 2[1-10] & 5.2[(-227)-11]  & 18.4[$\leq$32.8]\\
 \hline
 high-EW &   8.3[7.5-8.9] & 0.01[0.00-0.32] & 0.005[all] & 0.04[0.005-3] &   0.03[0.002-1.5] &  0.03[0.0002-65]  & 2[0.3-10] & 7.4[(-227)-13] & 14.4[$\leq$25.7] \\
 low-EW &  8.7[8.5-9.2] & 0.23[0.02-0.32] & -0.014 [all] & 0.02[0.007-3] &      0.007[0.003-0.85] & 50[0.03-250]  & 6[3-50] & -1.6[(-100)-9] & 7.1[$\leq$12.6]
 \enddata
\tablecomments{age$_{\rm SF}$, $\tau$ and t$^*$ are expressed in units of Gyr, the SFR estimations in M$_{\odot}$ yr$^{-1}$, M$_*$ in M$_{\odot}$ and E(B-V) in magnitudes. t$^*$ corresponds to the mean age of the stars inside the galaxy (Equation \ref{tstar}). SFR is derived from the star formation history (Equation \ref{sfrh}). 
$<$SFR$>_{100}$ is the average SFR over the last 100 Myr (Equation \ref{aSFR}). The $\chi^2$ reported in the last column corresponds to the best fit and to the 68\% confidence level.}
\label{tab:bestfitparameter}
\end{deluxetable}

\end{turnpage}
\clearpage
\begin{turnpage}
\begin{deluxetable*}{cccccccccr}
\tabletypesize{\tiny}
\tablecaption{Allowed parameter ranges for exponentially decreasing ($\tau>0$), increasing ($\tau<0$), CSF ($\tau=\infty$) models}
\tablewidth{0pt}
\tablehead{
\colhead{\bf{subsample}}      &   \colhead{\bf{log(M$^*$)}}       &
   \colhead{\bf{E(B-V)}}    &
     \colhead{\bf{$\tau$}}    & \colhead{\bf{age$_{\rm SF}$}}          &
\colhead{\bf{t$^*$}}        &
\colhead{\bf{SFR}} &  \colhead{\bf{$<$SFR$>_{100}$}} &  \colhead{\bf{age$_{\rm SF}$/$\tau$}} &\colhead{\bf{$\chi^2_{best}$ [$\chi^2_{68\%}$]}}}
\startdata
IRAC-bright &  9.3[9.1-9.6] & 0.39[0.30-0.48] & 0.005[all] & 0.007[0.006-0.022] &  0.005[0.003-0.011] & 130[40-600]  & 21[10-35] &  1.5[0.002-2.07]  & 12.0[$\leq$21.4]\\
&  9.4[9.2-9.7] & 0.40[0.30-0.50] & -0.687[all] & 0.009[0.006-0.404]  &  0.004[0.002-0.16] & 260[80-1300] & 23[10-120] &  -0.009[(-64)-(-0.002)] & 12[$\leq$21.4] \\
&  9.4[9.2-9.5] & 0.40[0.32-0.46] & $\infty$ & 0.009[0.006-0.015] &  0.004[0.002-0.007] &     260[110-530] & 23[15-30] & SFR$_{corr}$(UV)=130[70-210] &  12.0[$\leq$19.1] \\
IRAC-faint &  8.4[7.9-8.9] & 0.20[0.00-0.27] & 0.005[all] & 0.031[0.007-0.8] &    0.03[0.004-0.42] & 0.1[0.0003-30]  & 3[1-7] & 6.2[0.002-12.8]  & 21.8[$\leq$38.9]\\
&  8.6[8.0-9.0] & 0.07[0.00-0.30] & -0.118[all] & 1.70[0.009-3] &     0.1[0.003-1.5] &         3[1-80] & 2[1-15] & -14.5[(-220)-(-0.002)] & 23.5[$\leq$38.9] \\
&  8.3[8.0-8.9] & 0.11[0.00-0.25] & $\infty$ & 0.045[0.009-0.7] &     0.02[0.005-0.36] &   5[1-25] & 2[1-5] & SFR$_{corr}$(UV)=4[2-12] &  23.7[$\leq$34.7] \\
\hline
red-LAEs &  9.6[9.3-10.0] & 0.46[0.25-0.58] & 0.083[all] & 0.009[0.005-0.128] &   0.04[0.002-0.07] &    400[1-1595]  &   35[20-120] &   0.1[0.001-7.4]  & 8.0[$\leq$13.9]\\
&  9.7[9.3-10.0] & 0.48[0.32-0.59] & -0.005[all]  &   0.037[0.005-2.1] &    0.003[0.002-0.57] & 1090[70-2880] & 55[20-340] & -7.4[(-80)-(-0.001)] & 7.8[$\leq$13.9] \\
&   9.6[9.4-9.9] & 0.46[0.35-0.57] & $\infty$ & 0.009[0.005-0.08] &    0.004[0.002-0.04] &     400[100-1400] & 35[20-80] & SFR$_{corr}$(UV)=220[90-530] & 8.0[$\leq$12.4] \\
blue-LAEs &  8.4[8.3-9.0] & 0.18[0.00-0.23] & 0.014[all] & 0.01[0.009-0.202] &  0.007[0.005-0.1] & 20[0.02-30]  & 3[2-10] &  0.3[0.003-9]  & 6.3[$\leq$11.0]\\
&  8.5[8.4-8.9] & 0.17[0.05-0.25] & -0.014[all] & 0.227[0.01-3] &       0.01[0.005-0.85] &  25[4-110] & 3[2-20] &  -16[(-100)-(-0.003)] & 6.2[$\leq$11.0] \\
&  8.5[8.4-8.9] & 0.17[0.07-0.22] & $\infty$ & 0.02[0.01-0.16] &  0.009[0.005-0.08] &  16[5-30] & 3[2-6] & SFR$_{corr}$(UV)=10[5-16] &  6.5[$\leq$ 9.8] \\
\hline
  UV-bright &  9.0[8.9-9.4] & 0.29[0.09-0.35] & 0.007[all] & 0.009[0.007-0.064] &     0.006[0.004-0.05] &    50[1-150]  & 9[7-25] & 1.2[0.002-5.5]  & 9.5[$\leq$16.2]\\
&  9.1[8.8-9.2] & 0.32[0.23-0.38] &  -0.01[all] & 0.013[0.009-0.404] &    0.004[0.003-0.16] &  160[40-410] &  10[6-70] & -1.2[(-64)-(-0.002)] & 9.1[$\leq$16.2] \\
  &  9.0[8.9-9.1] & 0.30[0.25-0.34] & $\infty$ & 0.01[0.009-0.018] &    0.005[0.005-0.009] & 100[50-150] &      10[9-12] & SFR$_{corr}$(UV)=50[35-70] &  10.1[$\leq$14.5] \\
UV-faint &  8.3[7.9-9.0] & 0.04[0.00-0.23] & 0.005[all] & 0.03[0.009-0.806] &   0.02[0.005-0.43] &         0.2[0.002-18]  & 2[1-6] & 5.2[0.003-11]  & 18.4[$\leq$32.8]\\
&   8.6[7.9-9.0] & 0.04[0.00-0.24] & -1.39[all] & 0.286[0.01-3] &   0.14[0.005-1.5] &  2[1-64] &  2[1-10] & -0.2[(-227)-(-0.003)] & 19.5[$\leq$32.8] \\
&  8.6[8.0-8.9] & 0.03[0.00-0.21] & $\infty$ & 0.286[0.012-0.806] &      0.14[0.006-0.4] & 2[1-10] & 2[1-4] & SFR$_{corr}$(UV)=2[1-8] &  19.5[$\leq$29.3] \\
\hline
 high-EW & 8.3[7.5-8.8] &  0.01[0.00-0.30] & 0005[all]  & 0.04[0.005-0.806] &   0.03[0.002-0.4] & 0.03[0.0002-35] & 2[0.3-5] & 7.4[0.001-13] & 14.4[$\leq$25.7] \\
 &  8.1[7.5-8.9] & 0.11[0.00-0.32] & -0.687[all]  & 0.05[0.005-3] &       0.02[0.002-1.5] & 3[1-60] & 2[0.3-10] &  -0.06[(-227)-(-0.001)] & 14.8[$\leq$25.7] \\
&   8.1[7.6-8.7] & 0.11[0.00-0.27] & $\infty$ & 0.037[0.005-0.8] &     0.02[0.002-0.4] & 3[1-20] & 1[0.4-3] & SFR$_{corr}$(UV)=2[1-9] & 14.9[$\leq$22.9] \\ 
low-EW &  8.7[8.5-9.2] & 0.21[0.02-0.29] & 0.058[all] & 0.015[0.007-0.227] &     0.008[0.004-0.116] & 30[0.03-80] & 5[3-20] & 0.3[0.002-9] & 7.1[$\leq$12.7] \\
&  8.7[8.5-9.2] & 0.23[0.09-0.32] & -0.014[all] & 0.02[0.009-3] &  0.007[0.003-0.85] &  50[7-250] & 6[3-50] & -1.6[(-100)-(-0.002)] & 7.1[$\leq$12.7] \\
&  8.7[8.6-9.1] & 0.23[0.11-0.28] & $\infty$ & 0.013[0.009-0.161] &    0.006[0.005-0.08] & 40[9-70] & 5[4-14] & SFR$_{corr}$(UV)=25[10-40] & 7.2[$\leq$11.3] 
\enddata
\tablecomments{The units are the same as in Tab. \ref{tab:bestfitparameter}. For each sub-sample we report the SED fitting parameters for an exponentially decreasing, increasing and CSF ($\tau=18$ Gyr in our code) histories. In the case of a CSF model we report the SFR$_{corr}$(UV) parameter instead of the age$_{\rm SF}$/$\tau$. SFR$_{corr}$(UV) is also expressed in M$_{\odot}$yr$^{-1}$. To obtain dust-corrected SFR(UV) we apply the Calzetti (2000) dust law, using the E(B-V) allowed parameter ranges. As a reference the dust-uncorrected SFR(UV) is $3.9\pm0.5$ M$_{\odot}$yr$^{-1}$ for the FULLsample.}
\label{tab:last}
\end{deluxetable*}

\end{turnpage}

\clearpage

\begin{deluxetable*}{|c|c|c|c|c|c|c|c|c|}
\tabletypesize{\scriptsize}
\tablecaption{X-ray stacking results for the FULLsample}
\tablewidth{0pt}
\startdata
\hline
\bf{obs soft}          & \bf{obs hard}         &    \bf{rest hard} &      \bf{rest hard} & \bf{SFR$_{Persic}$} & \bf{SFR$_{Lehmer}$} & \bf{SFR$_{corr}$(UV)} & \bf{$<$SFR$>_{100}$} & \bf{SFR} \\
\bf{flux$_{X-ray}$} & \bf{flux$_{X-ray}$} & \bf{flux$_{X-ray}$} & \bf{L$_{X-ray}$} & & & & & \\
\hline
$\leq$2.2E-18 & (4.6$\pm$1.9)E-17 & (3.8$\pm$1.5)E-17 & (12$\pm$5)E+41 & 1180$\pm$480  & 730$\pm$300 & 15[6-30]     & 4[2-30] & 35[0.003-170] 
\enddata
\tablecomments{The units are erg s$^{-1}$ cm$^{-2}$ for the observed-frame soft-band (obs soft flux$_{X-ray}$), observed-frame hard-band (obs hard flux$_{X-ray}$), and rest-frame hard-band (obs hard flux$_{X-ray}$) X-ray flux, erg s$^{-1}$ for the rest-frame hard-band X-ray luminosity (rest hard L$_{X-ray}$), and M$_{\odot}$ yr$^{-1}$ for the star formation rate estimates. The dust-corrected SFR(UV) corresponds to the values calculated assuming a CSF model, while the instantaneous and averaged SFR considering the entire parameter space.}
\label{tab:Xray}
\end{deluxetable*}

\begin{thebibliography}{54}
\expandafter\ifx\csname natexlab\endcsname\relax\def\natexlab#1{#1}\fi

\bibitem[{{Acquaviva} {et~al.}(2011){Acquaviva}, {Gawiser}, \&
  {Guaita}}]{Acq2011}
{Acquaviva}, V., {Gawiser}, E., \& {Guaita}, L. 2011, ArXiv 1101.2215

\bibitem[{{Bertin} \& {Arnouts}(1996)}]{bertin1996}
{Bertin}, E. \& {Arnouts}, S. 1996, \aaps, 117, 393

\bibitem[{{Brinchmann} {et~al.}(2008){Brinchmann}, {Pettini}, \& {Charlot}}]{Brinchmann:2008}
{Brinchmann}, J., {Pettini}, M.,
  \& {Charlot}, S. 2008, \mnras, 385, 769

\bibitem[{{Bruzual}(2003)}]{BruzualCharlot2003}
{Bruzual}, G., \& {Charlot}, S. 2003, \mnras, 344, 1000  

\bibitem[{{Bond} {et~al.}(2009){Bond}, {Gawiser}, {Gronwall}, {Ciardullo},
  {Altmann}, \& {Schawinski}}]{Bond2009}
{Bond}, N.~A., {Gawiser}, E., {Gronwall}, C., {Ciardullo}, R., {Altmann}, M.,
  \& {Schawinski}, K. 2009, \apj, 705, 639
  
\bibitem[{{Bond} {et~al.}(2011){Bond},{Gawiser},{Guaita},{Padilla}, {Gronwall}, {Ciardullo}, \& {Lai}}]{Bond2011}{Bond}, N., {Gawiser}, E.,{Guaita}, L., {Padilla}, N., {Gronwall}, C., {Ciardullo}, R. \& {Lai}, K.
2011, ArXiv 1101.2215  

\bibitem[{{Calzetti} {et~al.}(2000){Calzetti}, {Armus}, {Bohlin}, {Kinney}, {Koornneef}, \& {Storchi-Bergmann}}]{Calzetti2000}
{Calzetti}, D., {Armus}, L., {Bohlin}, R.~C., {Kinney}, A.~L. {Koornneef}, J., {Storchi-Bergmann}, T. 2000, \apj, 533, 682

\bibitem[{{Cardamone} {et~al.}(2010){Cardamone}, {van Dokkum}, {Urry},
  {Taniguchi}, {Gawiser}, {Brammer}, {Taylor}, {Damen}, {Treister}, {Cobb},
  {Bond}, {Schawinski}, {Lira}, {Murayama}, {Saito}, \&
  {Sumikawa}}]{Cardamone:2010}
{Cardamone}, C.~N., {van Dokkum}, P.~G., {Urry}, C.~M., {Taniguchi}, Y.,
  {Gawiser}, E., {Brammer}, G., {Taylor}, E., {Damen}, M., {Treister}, E.,
  {Cobb}, B.~E., {Bond}, N., {Schawinski}, K., {Lira}, P., {Murayama}, T.,
  {Saito}, T., \& {Sumikawa}, K. 2010, \apjs, 189, 270

\bibitem[{{Chabrier}(2003)}]{chabrier2003}
{Chabrier}, G. 2003, \pasp, 115, 763

\bibitem[{{Cowie} \& {Hu}(1998)}]{CHu1998}
{Cowie}, L.~L. \& {Hu}, E.~M. 1998, \aj, 115, 1319

\bibitem[{{Damen}{ et~al.}(2011)}]{Damen2010}
{Damen}, M. 2011, \apj, 727, 1

\bibitem[{{Dunkley} {et~al.}(2009){Dunkley}, {Komatsu}, {Nolta}, {Spergel},
  {Larson}, {Hinshaw}, {Page}, {Bennett}, {Gold}, {Jarosik}, {Weiland},
  {Halpern}, {Hill}, {Kogut}, {Limon}, {Meyer}, {Tucker}, {Wollack}, \&
  {Wright}}]{Dunkley:2009}
{Dunkley}, J., {Komatsu}, E., {Nolta}, M.~R., {Spergel}, D.~N., {Larson}, D.,
  {Hinshaw}, G., {Page}, L., {Bennett}, C.~L., {Gold}, B., {Jarosik}, N.,
  {Weiland}, J.~L., {Halpern}, M., {Hill}, R.~S., {Kogut}, A., {Limon}, M.,
  {Meyer}, S.~S., {Tucker}, G.~S., {Wollack}, E., \& {Wright}, E.~L. 2009,
  \apjs, 180, 306

\bibitem[{{Erb} {et~al.}(2006{\natexlab{a}}){Erb}, {Steidel}, {Shapley},
  {Pettini}, {Reddy}, \& {Adelberger}}]{Erb2006b}
{Erb}, D.~K., {Steidel}, C.~C., {Shapley}, A.~E., {Pettini}, M., {Reddy},
  N.~A., \& {Adelberger}, K.~L. 2006{\natexlab{a}}, \apj, 647, 128

\bibitem[{{Erb} {et~al.}(2006{\natexlab{b}}){Erb}, {Steidel}, {Shapley},
  {Pettini}, {Reddy}, \& {Adelberger}}]{Erb:2006}
---. 2006{\natexlab{b}}, \apj, 646, 107

\bibitem[{{Finkelstein} {et~al.}(2008){Finkelstein}, {Rhoads}, {Malhotra},
  {Grogin}, \& {Wang}}]{Fin:2008}
{Finkelstein}, S.~L., {Rhoads}, J.~E., {Malhotra}, S., {Grogin}, N., \& {Wang},
  J. 2008, \apj, 678, 655

\bibitem[{{Francke} {et~al.}(2008){Francke}, {Gawiser}, {Lira}, {Treister},
  {Virani}, {Cardamone}, {Urry}, {van Dokkum}, \& {Quadri}}]{Francke:2008}
{Francke}, H., {Gawiser}, E., {Lira}, P., {Treister}, E., {Virani}, S.,
  {Cardamone}, C., {Urry}, C.~M., {van Dokkum}, P., \& {Quadri}, R. 2008,
  \apjl, 673, L13

\bibitem[{{Gawiser}(2009)}]{Gawiser2009}
{Gawiser}, E. 2009, New Astronomy Review, 53, 50

\bibitem[{{Gawiser} {et~al.}(2007){Gawiser}, {Francke}, {Lai}, {Schawinski},
  {Gronwall}, {Ciardullo}, {Quadri}, {Orsi}, {Barrientos}, {Blanc}, {Fazio},
  {Feldmeier}, {Huang}, {Infante}, {Lira}, {Padilla}, {Taylor}, {Treister},
  {Urry}, {van Dokkum}, \& {Virani}}]{Gawiser:2007}
{Gawiser}, E., {Francke}, H., {Lai}, K., {Schawinski}, K., {Gronwall}, C.,
  {Ciardullo}, R., {Quadri}, R., {Orsi}, A., {Barrientos}, L.~F., {Blanc},
  G.~A., {Fazio}, G., {Feldmeier}, J.~J., {Huang}, J.-S., {Infante}, L.,
  {Lira}, P., {Padilla}, N., {Taylor}, E.~N., {Treister}, E., {Urry}, C.~M.,
  {van Dokkum}, P.~G., \& {Virani}, S.~N. 2007, \apj, 671, 278

\bibitem[{{Gawiser} {et~al.}(2006{\natexlab{a}}){Gawiser}, {van Dokkum},
  {Gronwall}, {Ciardullo}, {Blanc}, {Castander}, {Feldmeier}, {Francke},
  {Franx}, {Haberzettl}, {Herrera}, {Hickey}, {Infante}, {Lira}, {Maza},
  {Quadri}, {Richardson}, {Schawinski}, {Schirmer}, {Taylor}, {Treister},
  {Urry}, \& {Virani}}]{Gawiser:2006b}
{Gawiser}, E., {van Dokkum}, P.~G., {Gronwall}, C., {Ciardullo}, R., {Blanc},
  G.~A., {Castander}, F.~J., {Feldmeier}, J., {Francke}, H., {Franx}, M.,
  {Haberzettl}, L., {Herrera}, D., {Hickey}, T., {Infante}, L., {Lira}, P.,
  {Maza}, J., {Quadri}, R., {Richardson}, A., {Schawinski}, K., {Schirmer}, M.,
  {Taylor}, E.~N., {Treister}, E., {Urry}, C.~M., \& {Virani}, S.~N.
  2006{\natexlab{a}}, \apjl, 642, L13

\bibitem[{{Gawiser} {et~al.}(2006{\natexlab{b}}){Gawiser}, {van Dokkum},
  {Herrera}, {Maza}, {Castander}, {Infante}, {Lira}, {Quadri}, {Toner},
  {Treister}, {Urry}, {Altmann}, {Assef}, {Christlein}, {Coppi}, {Dur{\'a}n},
  {Franx}, {Galaz}, {Huerta}, {Liu}, {L{\'o}pez}, {M{\'e}ndez}, {Moore},
  {Rubio}, {Ruiz}, {Toft}, \& {Yi}}]{Gawiser:2006a}
{Gawiser}, E., {van Dokkum}, P.~G., {Herrera}, D., {Maza}, J., {Castander},
  F.~J., {Infante}, L., {Lira}, P., {Quadri}, R., {Toner}, R., {Treister}, E.,
  {Urry}, C.~M., {Altmann}, M., {Assef}, R., {Christlein}, D., {Coppi}, P.~S.,
  {Dur{\'a}n}, M.~F., {Franx}, M., {Galaz}, G., {Huerta}, L., {Liu}, C.,
  {L{\'o}pez}, S., {M{\'e}ndez}, R., {Moore}, D.~C., {Rubio}, M., {Ruiz},
  M.~T., {Toft}, S., \& {Yi}, S.~K. 2006{\natexlab{b}}, \apjs, 162, 1
  
\bibitem[{{Ghosh}{ et~al.}(2001)}]{Ghosh2001}
{Ghosh}, P., \& {White}, N.~E. 2001, \apjl, 559, L97  

\bibitem[{{Grazian} {et~al.}(2007){Grazian}, {Salimbeni}, {Pentericci},
  {Fontana}, {Nonino}, {Vanzella}, {Cristiani}, {de Santis}, {Gallozzi},
  {Giallongo}, \& {Santini}}]{Grazian2007}
{Grazian}, A., {Salimbeni}, S., {Pentericci}, L., {Fontana}, A., {Nonino}, M.,
  {Vanzella}, E., {Cristiani}, S., {de Santis}, C., {Gallozzi}, S.,
  {Giallongo}, E., \& {Santini}, P. 2007, \aap, 465, 393

\bibitem[{{Gronwall} {et~al.}(2007){Gronwall}, {Ciardullo}, {Hickey},
  {Gawiser}, {Feldmeier}, {van Dokkum}, {Urry}, {Herrera}, {Lehmer}, {Infante},
  {Orsi}, {Marchesini}, {Blanc}, {Francke}, {Lira}, \&
  {Treister}}]{Gronwall:2007}
{Gronwall}, C., {Ciardullo}, R., {Hickey}, T., {Gawiser}, E., {Feldmeier},
  J.~J., {van Dokkum}, P.~G., {Urry}, C.~M., {Herrera}, D., {Lehmer}, B.~D.,
  {Infante}, L., {Orsi}, A., {Marchesini}, D., {Blanc}, G.~A., {Francke}, H.,
  {Lira}, P., \& {Treister}, E. 2007, \apj, 667, 79

\bibitem[{{Guaita} {et~al.}(2010){Guaita}, {Gawiser}, {Padilla}, {Francke},
  {Bond}, {Gronwall}, {Ciardullo}, {Feldmeier}, {Sinawa}, {Blanc}, \&
  {Virani}}]{Guaita2010}
{Guaita}, L., {Gawiser}, E., {Padilla}, N., {Francke}, H., {Bond}, N.~A.,
  {Gronwall}, C., {Ciardullo}, R., {Feldmeier}, J.~J., {Sinawa}, S., {Blanc},
  G.~A., \& {Virani}, S. 2010, \apj, 714, 255

\bibitem[{{Kannappan} \& {Gawiser}(2007)}]{KannappanGawiser07}
{Kannappan}, S.~J. \& {Gawiser}, E. 2007, \apjl, 657, L5

\bibitem[{{Kennicutt}(1998)}]{Kennicutt:1998}
{Kennicutt}, Jr., R.~C. 1998, \araa, 36, 189

\bibitem[{{Kornei} {et~al.}(2010){Kornei}, {Shapley}, {Erb}, {Steidel},
  {Reddy}, {Pettini}, \& {Bogosavljevi{\'c}}}]{kornei2010}
{Kornei}, K.~A., {Shapley}, A.~E., {Erb}, D.~K., {Steidel}, C.~C., {Reddy},
  N.~A., {Pettini}, M., \& {Bogosavljevi{\'c}}, M. 2010, \apj, 711, 693
  
\bibitem[{{Kurczynski}{et~al.}(2010)}]{Kurczynski2010}
{Kurczynski}, P. 2010, ArXiv 1010.0290  

\bibitem[{{Lai} {et~al.}(2008){Lai}, {Huang}, {Fazio}, {Gawiser}, {Ciardullo},
  {Damen}, {Franx}, {Gronwall}, {Labbe}, {Magdis}, \& {van Dokkum}}]{Lai:2008}
{Lai}, K., {Huang}, J.-S., {Fazio}, G., {Gawiser}, E., {Ciardullo}, R.,
  {Damen}, M., {Franx}, M., {Gronwall}, C., {Labbe}, I., {Magdis}, G., \& {van
  Dokkum}, P. 2008, \apj, 674, 70

\bibitem[{{Lee} {et~al.}(2010){Lee}, {Ferguson}, {Somerville}, {Wiklind}, \&
  {Giavalisco}}]{Lee2010}
{Lee}, S., {Ferguson}, H.~C., {Somerville}, R.~S., {Wiklind}, T., \&
  {Giavalisco}, M. 2010, \apj, 725, 1644

\bibitem[{{Lehmer} {et~al.}(2010){Lehmer}, {Alexander}, {Bauer}, {Brandt}, {Goulding}, {Jenkins},{Ptak}, \& {Roberts}}]{Lehmer2010}{Lehmer}, B.~D., {Alexander}, D.~M., {Bauer}, F.~E., {Brandt}, W.~N., {Goulding}, A.~D., {Jenkins}, L.~P., {Ptak}, A., {Roberts}, T.~P. 2010, \apj, 724, 559

\bibitem[{{Madau}(1995)}]{Madau:1995}
{Madau}, P. 1995, \apj, 441, 18

\bibitem[{{Malhotra} \& {Rhoads}(2002)}]{MR:2002}
{Malhotra}, S. \& {Rhoads}, J.~E. 2002, \apjl, 565, L71

\bibitem[{{Maraston} {et~al.}(2010){Maraston}, {Pforr}, {Renzini}, {Daddi},
  {Dickinson}, {Cimatti}, \& {Tonini}}]{Maraston2010}
{Maraston}, C., {Pforr}, J., {Renzini}, A., {Daddi}, E., {Dickinson}, M.,
  {Cimatti}, A., \& {Tonini}, C. 2010, \mnras, 407, 830

\bibitem[{{Marigo} \& {Girardi}(2007)}]{MarigoGirardi2007}
{Marigo}, P. \& {Girardi}, L. 2007, \aap, 469, 239

\bibitem[{{Meurer} {et~al.}(1999){Meurer}, {Heckman}, \&
  {Calzetti}}]{Meurer1999}
{Meurer}, G.~R., {Heckman}, T.~M., \& {Calzetti}, D. 1999, \apj, 521, 64

\bibitem[{{Neufeld}(1991)}]{neufeld:1991}
{Neufeld}, D.~A. 1991, \apjl, 370, L85

\bibitem[{{Nilsson} {et~al.}(2007){Nilsson}, {M{\o}ller}, {M{\"o}ller},
  {Fynbo}, {Micha{\l}owski}, {Watson}, {Ledoux}, {Rosati}, {Pedersen}, \&
  {Grove}}]{Nilsson:2007}
{Nilsson}, K.~K., {M{\o}ller}, P., {M{\"o}ller}, O., {Fynbo}, J.~P.~U.,
  {Micha{\l}owski}, M.~J., {Watson}, D., {Ledoux}, C., {Rosati}, P.,
  {Pedersen}, K., \& {Grove}, L.~F. 2007, \aap, 471, 71

\bibitem[{{Nilsson} {et~al.}(2011){Nilsson}, {{\"O}stlin}, {M{\o}ller},
  {M{\"o}ller-Nilsson}, {Tapken}, {Freudling}, \& {Fynbo}}]{Nilsson2010}
{Nilsson}, K.~K., {{\"O}stlin}, G., {M{\o}ller}, P., {M{\"o}ller-Nilsson}, O., {Tapken}, C., {Freudling}, W., \& {Fynbo}, J.~P.~U. 2011, \aap, 529, A9

\bibitem[{{Nilsson} {et~al.}(2009){Nilsson}, {Tapken}, {M{\o}ller},
  {Freudling}, {Fynbo}, {Meisenheimer}, {Laursen}, \&
  {{\"O}stlin}}]{Nilsson:2009}
{Nilsson}, K.~K., {Tapken}, C., {M{\o}ller}, P., {Freudling}, W., {Fynbo},
  J.~P.~U., {Meisenheimer}, K., {Laursen}, P., \& {{\"O}stlin}, G. 2009, \aap,
  498, 13

\bibitem[{{Ono} {et~al.}(2010{\natexlab{a}}){Ono}, {Ouchi}, {Shimasaku},
  {Akiyama}, {Dunlop}, {Farrah}, {Lee}, {McLure}, {Okamura}, \&
  {Yoshida}}]{Ono2010}
{Ono}, Y., {Ouchi}, M., {Shimasaku}, K., {Akiyama}, M., {Dunlop}, J., {Farrah},
  D., {Lee}, J.~C., {McLure}, R., {Okamura}, S., \& {Yoshida}, M.
  2010{\natexlab{a}}, MNRAS, 402, 1580

\bibitem[{{Ono} {et~al.}(2010{\natexlab{b}}){Ono}, {Ouchi}, {Shimasaku},
  {Dunlop}, {Farrah}, {McLure}, \& {Okamura}}]{Ono2010b}
{Ono}, Y., {Ouchi}, M., {Shimasaku}, K., {Dunlop}, J., {Farrah}, D., {McLure}, R., \& {Okamura}, S. 2010{\natexlab{b}}, \apj, 724, 1524

\bibitem[{{Ouchi} {et~al.}(2008){Ouchi}, {Shimasaku}, {Akiyama}, {Simpson},
  {Saito}, {Ueda}, {Furusawa}, {Sekiguchi}, {Yamada}, {Kodama}, {Kashikawa},
  {Okamura}, {Iye}, {Takata}, {Yoshida}, \& {Yoshida}}]{Ouchi:2008}
{Ouchi}, M., {Shimasaku}, K., {Akiyama}, M., {Simpson}, C., {Saito}, T.,
  {Ueda}, Y., {Furusawa}, H., {Sekiguchi}, K., {Yamada}, T., {Kodama}, T.,
  {Kashikawa}, N., {Okamura}, S., {Iye}, M., {Takata}, T., {Yoshida}, M., \&
  {Yoshida}, M. 2008, \apjs, 176, 301

\bibitem[{{Ouchi} {et~al.}(2003){Ouchi}, {Shimasaku}, {Furusawa}, {Miyazaki},
  {Doi}, {Hamabe}, {Hayashino}, {Kimura}, {Kodaira}, {Komiyama}, {Matsuda},
  {Miyazaki}, {Nakata}, {Okamura}, {Sekiguchi}, {Shioya}, {Tamura},
  {Taniguchi}, {Yagi}, \& {Yasuda}}]{Ouchi:2003}
{Ouchi}, M., {Shimasaku}, K., {Furusawa}, H., {Miyazaki}, M., {Doi}, M.,
  {Hamabe}, M., {Hayashino}, T., {Kimura}, M., {Kodaira}, K., {Komiyama}, Y.,
  {Matsuda}, Y., {Miyazaki}, S., {Nakata}, F., {Okamura}, S., {Sekiguchi}, M.,
  {Shioya}, Y., {Tamura}, H., {Taniguchi}, Y., {Yagi}, M., \& {Yasuda}, N.
  2003, \apj, 582, 60

\bibitem[{{Ouchi} {et~al.}(2010){Ouchi}, {Shimasaku}, {Furusawa}, {SAITO},
  {Yoshida}, {Akiyama}, {Ono}, {Yamada}, {Ota}, {Kashikawa}, {Iye}, {Kodama},  {Okamura}, {Simpson}, \& {Yoshida}}]{Ouchi:2010}
{Ouchi}, M., {Shimasaku}, K., {Furusawa}, H., {SAITO}, T., {Yoshida}, M.,
  {Akiyama}, M., {Ono}, Y., {Yamada}, T., {Ota}, K., {Kashikawa}, N., {Iye},  M., {Kodama}, T., {Okamura}, S., {Simpson}, C., \& {Yoshida}, M. 2010, \apj, 723, 869

\bibitem[{{Papovich} {et~al.}(2001){Papovich}, {Dickinson}, \&
  {Ferguson}}]{Papovich2001}
{Papovich}, C., {Dickinson}, M., \& {Ferguson}, H.~C. 2001, \apjl, 559, 620

\bibitem[{{Papovich} {et~al.}(2011){Papovich}, {Finkelstein}, {Ferguson},
  {Lotz}, \& {Giavalisco}}]{Papovich2010}
{Papovich}, C., {Finkelstein}, S.~L., {Ferguson}, H.~C., {Lotz}, J.~M., \&
  {Giavalisco}, M. 2011, \mnras, 412, 1123

\bibitem[{{Pascarelle} {et~al.}(1998){Pascarelle}, {Windhorst}, \&
  {Keel}}]{Pascarelle1998}
{Pascarelle}, S.~M., {Windhorst}, R.~A., \& {Keel}, W.~C. 1998, \aj, 116, 2659

\bibitem[{{Pentericci} {et~al.}(2009){Pentericci}, {Grazian}, {Fontana},
  {Castellano}, {Giallongo}, {Salimbeni}, \& {Santini}}]{Pentericci:2009}
{Pentericci}, L., {Grazian}, A., {Fontana}, A., {Castellano}, M., {Giallongo},
  E., {Salimbeni}, S., \& {Santini}, P. 2009, \aap, 494, 553

\bibitem[{{Pentericci} {et~al.}(2007){Pentericci}, {Grazian}, {Fontana},
  {Salimbeni}, {Santini}, {de Santis}, {Gallozzi}, \&
  {Giallongo}}]{Pentericci:2007}
{Pentericci}, L., {Grazian}, A., {Fontana}, A., {Salimbeni}, S., {Santini}, P.,
  {de Santis}, C., {Gallozzi}, S., \& {Giallongo}, E. 2007, \aap, 471, 433

\bibitem[{{Persic} \& {Rephaeli}(2002)}]{Persic2002}
{Persic}, M. \& {Rephaeli}, Y. 2002, \aap, 382, 843

\bibitem[{{Persic}{~ et~al.}(2004)}]{Persic2004}
{Persic}, M. 2004, \aap, 419, 849

\bibitem[{{Pirzkal} {et~al.}(2007){Pirzkal}, {Malhotra}, {Rhoads}, \&
  {Xu}}]{Pirzkal:2007}
{Pirzkal}, N., {Malhotra}, S., {Rhoads}, J.~E., \& {Xu}, C. 2007, \apj, 667, 49 

\bibitem[{{Salpeter}(1955)}]{Salpeter1955}
{Salpeter}, E.~E. 1955, \apj, 121, 161

\bibitem[{{Schaerer} \& {Vacca}(1998)}]{ScVacca:1998}
{Schaerer}, D. \& {Vacca}, W.~D. 1998, \apj, 497, 618

\bibitem[{{Shapley} {et~al.}(2001){Shapley}, {Steidel}, {Adelberger},
  {Dickinson}, {Giavalisco}, \& {Pettini}}]{Shapley:2001}
{Shapley}, A.~E., {Steidel}, C.~C., {Adelberger}, K.~L., {Dickinson}, M.,
  {Giavalisco}, M., \& {Pettini}, M. 2001, \apj, 562, 95

\bibitem[{{Sheth} \& {Tormen}(1999)}]{Shetormen:1999}
{Sheth}, R.~K. \& {Tormen}, G. 1999, \mnras, 308, 119

\bibitem[{{Steidel} {et~al.}(2004){Steidel}, {Shapley}, {Pettini},
  {Adelberger}, {Erb}, {Reddy}, \& {Hunt}}]{Steidel:2004}
{Steidel}, C.~C., {Shapley}, A.~E., {Pettini}, M., {Adelberger}, K.~L., {Erb},
  D.~K., {Reddy}, N.~A., \& {Hunt}, M.~P. 2004, \apj, 604, 534

\bibitem[{{Tapken} {et~al.}(2007){Tapken}, {Appenzeller}, {Noll}, {Richling},
  {Heidt}, {Meink{\"o}hn}, \& {Mehlert}}]{Tapken:2007}
{Tapken}, C., {Appenzeller}, I., {Noll}, S., {Richling}, S., {Heidt}, J.,
  {Meink{\"o}hn}, E., \& {Mehlert}, D. 2007, \aap, 467, 63

\bibitem[{{Taylor} {et~al.}(2009){Taylor}, {Franx}, {van Dokkum}, {Quadri},
  {Gawiser}, {Bell}, {Barrientos}, {Blanc}, {Castander}, {Damen},
  {Gonzalez-Perez}, {Hall}, {Herrera}, {Hildebrandt}, {Kriek}, {Labb{\'e}},
  {Lira}, {Maza}, {Rudnick}, {Treister}, {Urry}, {Willis}, \&
  {Wuyts}}]{Taylor:2009}
{Taylor}, E.~N., {Franx}, M., {van Dokkum}, P.~G., {Quadri}, R.~F., {Gawiser},
  E., {Bell}, E.~F., {Barrientos}, L.~F., {Blanc}, G.~A., {Castander}, F.~J.,
  {Damen}, M., {Gonzalez-Perez}, V., {Hall}, P.~B., {Herrera}, D.,
  {Hildebrandt}, H., {Kriek}, M., {Labb{\'e}}, I., {Lira}, P., {Maza}, J.,
  {Rudnick}, G., {Treister}, E., {Urry}, C.~M., {Willis}, J.~P., \& {Wuyts}, S.
  2009, \apjs, 183, 295
  
\bibitem[{{Treister} {et~al.}(2009){Treister}, {Virani}, {Gawiser},
  {Urry}, {Lira}, {Francke}, {Blanc}, {Cardamone}, {Damen}, {Taylor}, \& {Schawinski}}]{Treister:2009}
{Treister}, E. and {Virani}, S. and {Gawiser}, E. and {Urry}, C.~M. and {Lira}, P. and {Francke}, H. and {Blanc}, G.~A. and {Cardamone}, C.~N. and {Damen}, M. and {Taylor}, E.~N. \& {Schawinski}, K.  2009, \apj, 693, 1713  

\bibitem[{{Verhamme} {et~al.}(2006){Verhamme}, {Schaerer}, \&
  {Maselli}}]{V:2006}
{Verhamme}, A., {Schaerer}, D., \& {Maselli}, A. 2006, \aap, 460, 397

\bibitem[{{Walcher} {et~al.}(2011){Walcher}, {Groves}, {Budavari}, \&
  {Dale}}]{Walcher2010} {Walcher}, C.~J., {Groves}, B., {Budavari}, T., \& {Dale}, D. 2011, \apss, 331,1 
  
\bibitem[{{Wang} {et~al.}(2004){Wang}, {Malhotra}, {Rhoads}, \&
  {Norman},}]{Wang:2004}
{Wang}, J.~X. and {Malhotra}, S. and {Rhoads}, J.~E. \& {Norman}, C.~A. 2004, \apjl, 612, L109

\bibitem[{{Yuma} {et~al.}(2010){Yuma}, {Ohta}, {Yabe}, {Shimasaku}, {Yoshida},
  {Ouchi}, {Iwata}, \& {Sawicki}}]{Yuma:2010}
{Yuma}, S., {Ohta}, K., {Yabe}, K., {Shimasaku}, K., {Yoshida}, M., {Ouchi},
  M., {Iwata}, I., \& {Sawicki}, M. 2010, \apj, 720, 1016

\end{thebibliography}
\end{document}